\documentclass[natbib]{svjour3}                     
\smartqed  
 \usepackage{aps-bibstyle}  
\usepackage{graphicx}
\usepackage{color}
\usepackage{ulem}
\begin{document}
\title{Outgassing History and Escape of the Martian Atmosphere and Water Inventory}

\titlerunning{Outgassing and loss of Mars' atmosphere}

\author{Helmut Lammer \and
Eric Chassefi\`{e}re \and
\"{O}zg\"{u}r Karatekin \and
Achim Morschhauser \and
Paul B. Niles \and
Olivier Mousis \and
Petra Odert \and
Ute V. M\"{o}stl \and
Doris Breuer \and
V\'{e}ronique Dehant \and
Matthias Grott \and
Hannes Gr\"{o}ller \and
Ernst Hauber \and
L\^{e} Binh San Pham}

\authorrunning{H. Lammer et al.}

\institute{H. Lammer, H. Gr\"{o}ller, P. Odert\at
Space Research Institute, Austrian Academy of Sciences, Schmiedlstr. 6, A-8042 Graz, Austria\\
\email{helmut.lammer@oeaw.ac.at}
\and
E. Chassefi\`{e}re \at
Univ. Paris-Sud, Laboratoire IDES, CNRS, UMR8148, Orsay, F-91405, France\\
\and
V. Dehant, \"{O}. Karatekin, L.B.S. Pham \at
Royal Observatory of Belgium, Brussels, Belgium\\
\and
D. Breuer, M. Grott, E. Hauber, A. Morschhauser \at
German Aerospace Center, Institute of Planetary Research, Rutherfordstr. 2, D-12489 Berlin, Germany\\
\and
P.B. Niles \at
Astromaterials Research and Exploration Science Johnson Space Center, NASA, Houston Texas, USA\\
\and
O. Mousis \at
Observatoire de Besan\c{c}on, 41 bis, avenue de l'Observatoire B.P. 1615 avenue de l'Observatoire,
F-25010 Besan\c{c}on, France,\\
and Universit\'e de Toulouse; UPS-OMP; CNRS-INSU; IRAP; 14 Avenue Edouard Belin, 31400 Toulouse, France\\
\and
U.V. M\"{o}stl, P. Odert \at
Institute for Physics/IGAM, University of Graz, Universit\"{a}tsplatz 5, A-8010, Graz, Austria}

\date{Received: date / Accepted: date}
\maketitle
\begin{abstract}
The evolution and escape of the martian atmosphere and the planet's water inventory
can be separated into an early and late evolutionary epoch. The first epoch started
from the planet's origin and lasted $\sim$500 Myr. Because of the high EUV flux of the young Sun and Mars'
low gravity it was accompanied by hydrodynamic blow-off of
hydrogen and strong thermal escape rates of dragged heavier species such as O and C atoms.
After the main part of the protoatmosphere was lost, impact-related volatiles and mantle outgassing may have resulted in accumulation of a secondary CO$_2$ atmosphere of a few tens to a few hundred mbar around $\sim$4--4.3 Gyr ago. The evolution of the atmospheric surface pressure and water inventory of such a secondary atmosphere during the second epoch which lasted from the end of the Noachian until today was most likely determined by a complex interplay of various nonthermal atmospheric escape processes, impacts, carbonate precipitation, and serpentinization during the Hesperian and Amazonian epochs which led to the present day surface pressure.
\end{abstract}
\section{Introduction}
The present martian atmosphere is the result of numerous interacting processes.
On the one hand, these include atmospheric sinks such as erosion by impacts, thermal and nonthermal escape,
extreme ultraviolet (EUV) radiation, as well as solar wind forcing (e.g. Lundin et al. 2007).
On the other hand, atmospheric sources such as volcanic outgassing or delivery of volatiles by impacts have also to be
taken into account for understanding atmospheric evolution. Furthermore, the atmosphere can interact
with crustal reservoirs by CO$_2$ weathering and hydration processes, which occur at the surface and/or in the crust
(e.g., Zent and Quinn 1995; Bandfield et al. 2003; Becker et al. 2003; Lundin et al.
2007; Lammer et al. 2008; Pham et al. 2009; Tian et al. 2009; Philips et al. 2010).
A schematic synopsis of these interactions is presented in Fig. 1.
\begin{figure}[h]
\begin{center}
\includegraphics[width=0.85\columnwidth]{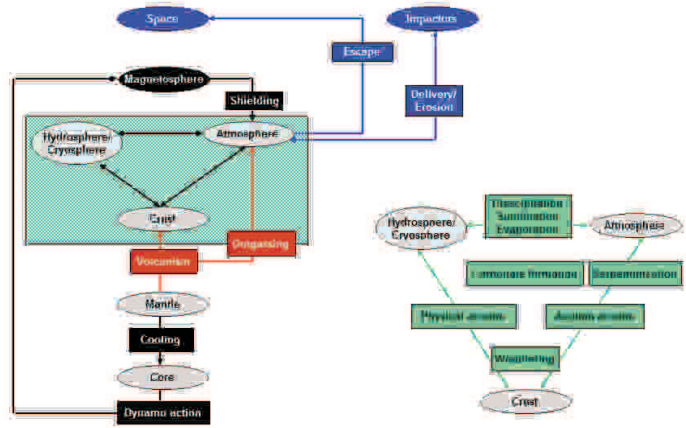}
\caption{Sketch showing important interactions between the main reservoirs, i.e. atmosphere, hydrosphere/cryosphere, crust, mantle and core (ovals in light blue), that have been addressed in the present work. Volcanism results in the formation of the crust and the associated degassing of the mantle produces an atmosphere with time (red boxes). Dynamo action in the core, which is triggered by efficient heat transfer in the mantle, and the subsequent shielding of the atmosphere prevents or reduces atmospheric erosion by non-thermal processes (black boxes). The erosion of the atmosphere to space can be caused by solar influx or by impacts (blue boxes). The latter may also deliver volatiles to the atmosphere depending on the impactors size and composition. The complex interactions between atmosphere, hydrosphere/cryosphere and crust (green dotted area) are shown in more detail on the right side (green boxes).}
\end{center}
\end{figure}

The aim of this work is to review the latest knowledge on
the evolution of the martian atmosphere since the planet's origin $\sim$4.55 Gyr ago.
In Sect. 2 we discuss the delivery of volatiles, the planet's early hydrogen-rich protoatmosphere,
and point out possible reasons why there is an apparent deficiency of noble gases in the present atmosphere.
In Sect. 3 we discuss different approaches to constrain volcanic outgassing rates of CO$_2$ and H$_2$O. In Sect. 4 we consider the
role of atmospheric impact erosion and delivery in the early martian environment.
In Sect. 5, we discuss the efficiency of EUV-powered escape during the early Noachian and its influence on the growth of a secondary CO$_2$ atmosphere.
In Sect. 6 we briefly address consequences of the late heavy bombardment (LHB) on the martian atmosphere and its climate, $\sim$3.7--4 Gyr ago. Finally,
Sects. 7 and 8 focus on nonthermal atmospheric escape to space and on possible surface sinks of CO$_2$ and H$_2$O allowing the surface pressure to reach its present-day value.
\section{Origin and Delivery of Volatiles to Mars}
The sources and evolutionary histories of volatiles composing the martian atmosphere are poorly understood. They are related to
the sources that delivered significant amounts of water to early Mars, which have implications for the formation of the planet's protoatmosphere.
Furthermore, isotope variations in volatiles have the potential to provide insights into the origin and atmosphere modification processes in terrestrial planets,
possibly related to the observation that the noble gases appear strongly depleted in the martian atmosphere
compared to those of Earth and Venus.
\subsection{Water Delivery and Formation of the Martian Protoatmosphere}
Four main processes are responsible for the early formation of an atmosphere:
\begin{itemize}
\item capture and accumulation of gasses from the planetary nebula,
\item catastrophic outgassing due to magma ocean solidification,
\item impacts,

and

\item later degassing by volcanic processes.
\end{itemize}
As long as nebula gas is present, growing protoplanets can capture hydrogen and He which form gaseous
envelopes around the rocky core (e.g., Hayashi et al. 1979; Rafikov 2006). Depending on the
host star's radiation and plasma outflow, the nebula dissipation time, the planet's orbital location and
the number and orbital location of additional planets in the system, according to Hayashi et al. (1979),
planetary embryos with the mass of $\sim$0.1$M_{\rm Earth}$ can capture hydrogen and other nebula gas from the nebula during
$\sim$3 Myr with an equivalent amount of up to $\sim$55 times the hydrogen which is, present in the Earth's present day ocean.
Furthermore, noble gases delivered by comets accreted during this period were mixed with volatiles
remaining after an episode of strong atmospheric escape.

The initial water inventory of a planet is acquired from colliding planetesimals, growing planetary embryos,
impacting asteroids and comets (e.g., Lunine et al. 2003; Brasser 2012). Lunine et al. (2003) estimated the cumulative collision
probability between small bodies and Mars and found that Mars' initial water inventory may have equivalent
to $\sim$0.06--0.27 times that of an Earth ocean (EO), corresponding to a martian surface pressure of
$\sim$10--100 bar. Other simulations which considered different impact regimes suggest that Mars could
also have been drier (Horner 2009).
In a more recent study, Walsh et al. (2011) argues that the small mass of Mars indicates that the terrestrial planets in the Solar System
have formed from a narrow material annulus, rather than a disc extending to Jupiter. In such a scenario the truncation
of the outer edge of the disc was the result of the migration of the gas giants, which kept the martian mass small.
From cosmochemical constraints one can argue that Mars formed in a couple of Myr and can be considered in agreement
with the latest dynamical models as a planetary embryo that never grew to a real planet. In such a case most of
Mars' materials consisted of building blocks that formed in a region at $\sim$2--3 AU, and therefore, were more H$_2$O-rich
compared to the materials which formed Earth and Venus. From these arguments Brasser (2012) suggests that
Mars may have consisted of $\sim$0.1--0.2 wt.\% of water.

A substantial part of the initial inventory of volatiles could have
been outgassed as a consequence of the solidifcation of an early magma ocean (Elkins-Tanton 2008).
Water and carbon dioxide enter solidifying minerals in only small quantities and
are enriched in magma ocean liquids as solidification proceeds. Close to the surface at low pressure
these volatiles degas into the growing atmosphere.
Depending on the initial water/volatile content, which was built-in the planetary body during its growth
and the depth of the possible magma ocean $z_{\rm mag}$, steam atmospheres with a surface pressure between $\sim$30 (0.05 wt.\% H$_2$O, 0.01 wt.\% CO$_2$,
$z_{\rm mag}$$\sim$500 km) to $\sim$800 bar (0.5 wt.\% H$_2$O, 0.1 wt.\% CO$_2$, $z_{\rm mag}$$\sim$2000 km) (see Table 3, Elkins-Tanton 2008) could have
been catastrophically outgassed.  If early Mars
consisted of $\sim$0.1--0.2 wt.\% water (Brasser 2012) then a steam atmosphere with a surface pressure of
more than $\sim$60 bar could have been catastrophically outgassed (Elkins-Tanton 2008).
Although it is assumend that most volatiles are degassed into the early
atmosphere, a geodynamically significant quantity is still sequestered in
the solid cumulates. The amount is estimated to be as much as 750 ppm by weight OH for an initial water content of 0.5 wt.\%, and a minimum of 10 ppm by weight in
the driest cumulates of models beginning with just 0.05 wt.\% water (Elkins-Tanton 2008).
Even more water in the martian interior can be expected after the magma ocean solidification phase in the case of a shallow magma ocean in particular in the deep unmolden primordial mantle. In any case, these small water contents significantly lower the viscosity and possibly the
melting temperature of mantle materials, facilitating later volcanism,
as discussed below.

The early steam atmosphere could have remained stable for a few tens of Myr. During this early stage,
environmental conditions were determined by a high surface temperature and frequent impacts,
which could have reached up to $\sim$1500 K due to thermal blanketing and frequent impacts (e.g., Matsui and Abe 1986).
If such a steam atmosphere is not lost upon cooling,
the remaining H$_2$O vapor can condense and produces liquid water on the surface
or ice in case of a cold climate (e.g., Chassefi\`{e}re 1996).

If Mars originated with $\sim$0.1--0.2 wt\% H$_2$O, as long as the planet was surrounded by a captured dense nebula-based
hydrogen envelope, magma ocean related outgassed greenhouse gases (H$_2$O, CO$_2$, CH$_4$, NH$_3$)
would have been protected against dissociation because these heavy molecules would remain closer to the planet's surface compared to the
lighter hydrogen in the upper atmosphere. Depending on the amount and the lifetime of accumulated nebula gas and its evaporation time,
a combination of a possible H$_2$ greenhouse (Pierrehumbert and Gaidos 2011; Wordsworth 2012) and the outgassed greenhouse gases
may have provided warm and wet conditions on the martian surface for a few tens of Myr.

Finally, it is important to note that the previous investigations of planetary formation suffer from several unknowns including the sources of impactors across the inner Solar System.
Such work would require far more detailed model populations for the cometary and asteroidal sources, and would have to include a study of the effects of Oort cloud comets.
Because the results of Lunine et al. (2003) and Brasser (2012) are different from those of Horner et al. (2009), it is obvious that our knowledge of terrestrial planet formation and
hydration is currently insufficient because it is not possible to predict the real initial deuteration level on each of the planets considered.
This piece of evidence, combined with the fact that the D/H ratio in H$_2$O in comets is not homogeneous (Hartogh et al. 2011), indicate that the
water delivery mechanisms to the terrestrial planets can only be established within an uncertainty range.

\subsection{The Apparent Noble Gas Deficiency of the Martian Atmosphere}
The difference between the measured atmospheric abundances of non-radiogenic noble gases in Venus, Earth, and Mars is striking. It is well known that these abundances decline dramatically as one moves outward from Venus to Mars within the inner Solar System, with these two planets differing in abundance by up to two orders of magnitude (see Fig. 2). Therefore, understanding this variation is a key issue in understanding how the initial atmospheres of the terrestrial planets evolved to their current composition, and requires
to study the different delivery mechanisms of the volatiles accreted by these planets
(Pepin 1991; 2006; Owen et al. 1992; Owen and Bar-Nun 1995; Dauphas 2003; Marty and Meibom 2007).

In this context, recent $n$-body simulations have been performed by Horner et al. (2009) in order to study the impact rates experienced by the terrestrial planets as a result of diverse populations of potential impactors. These authors considered a wide range of plausible planetary formation scenarios for the terrestrial planets, and found that the different impact regimes experienced by Venus, Earth, and Mars could have resulted in significant differences between their individual hydration states over the course of their formation and evolution. Horner et al. (2009) found that, on average, the Earth most likely received a flux of impacting comets which is $\sim$3.4 times higher than that experienced by Mars. Assuming that the mass of noble gases delivered by comets to the terrestrial planets is proportional to the rate at which they impacted upon them, it is possible to derive $X_{\rm E} / X_{\rm M}$ (the ratio of the noble gas abundances (as a fraction of the total mass of the planet) between Earth and Mars) from the ratio of the number of comets impacting upon those two planets $N_E / N_M$ ($\sim$3.4), through the following relation (Mousis et al. 2010)
\begin{equation}
\frac{X_{\rm E}}{X_{\rm M}} = \frac{N_{\rm E}}{N_{\rm M}} \frac{M_{\rm M}}{M_{\rm E}},
\end{equation}
\noindent where $M_{\rm E}$ and $M_{\rm M}$ are the masses of Earth and Mars, respectively. From this relation, one can infer that the average noble gas abundance on Earth should be $\sim$0.37 times the martian noble gas abundances if these volatiles were solely delivered by comets. This result differs significantly from that inferred from measurements of noble gas abundances, which are observed to be approximately two orders of magnitude larger for the Earth compared to Mars. As a result, subsequent processes that occurred preferably during the post-impact period of Mars are required in order to explain its present-day atmospheric composition.
\begin{figure}
\begin{center}
\includegraphics[width=0.55\textwidth]{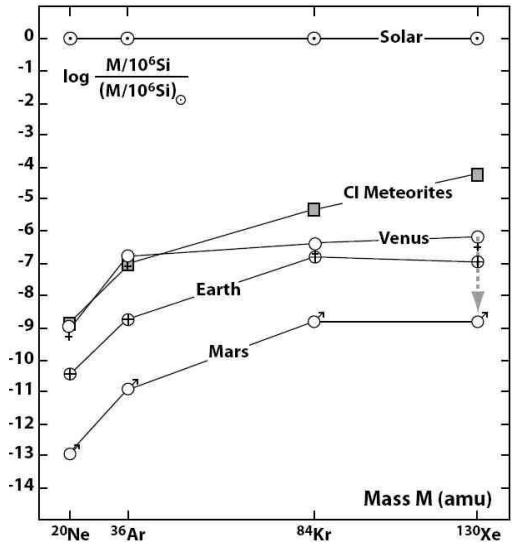}
\caption{Measured abundances of Ne, Ar, Kr, and Xe in the atmospheres of the terrestrial planets and primitive CI meteorites. The values shown for these gases are presented relative to their solar abundances, in units of atoms per 10$^6$ Si atoms (adapted from Fig. 2 of Pepin 1991). The vertical arrow pointing down indicates that the Venus atmospheric abundance of Xe is only an upper limit.}
\end{center}
\label{NGpepin}
\end{figure}
It has been proposed that atmospheric escape could have strongly altered the composition of the atmospheres of terrestrial planets (Pepin 1991; 1997; Dauphas 2003; Jakosky et al. 1994; Chassefi{\`e}re and Leblanc 2004). This hypothesis is supported by both Mars (SNC meteorites) and Earth, which show substantial fractionation of Xe isotopes compared to the plausible primitive sources of noble gases, i.e., solar wind (SW--Xe), meteorites (Q--Xe), or the hypothetical U--Xe source (Pepin 2006). This fractionation then suggests important losses of Xe and other noble gases from the early atmospheres of the Earth and Mars. Impact related loss processes might have been more important for the Earth and Mars than Venus because the latter planet would have escaped impacts of the magnitude that formed the Moon (Canup and Asphaug 2001) or created the largest basins on Mars (Andrews-Hanna et al. 2008). In the case of the Earth, the noble gas fractionation episode could have also been driven by impacts (Pepin 1991; Pepin 1997; Dauphas 2003) in combination with the high
EUV radiation of the young Sun (e.g., Ribas es al. 2005; Lammer et al. 2008).

Thus, in the case of Mars, the combination of impacts, EUV-powered hydrodynamic escape, planetary degassing, and fractionation by nonthermal atmospheric escape processes (Jakosky et al. 1994; Luhmann et al. 1992; Carr 1999; Chassefi{\`e}re and Leblanc 2004; Lammer et al. 2008) might have played an important role in sculpting the pattern of the noble gas abundances observed today.
An alternative hypothesis proposed to explain the Kr and Xe abundance differences between Earth and Mars is the presence of large amounts of CO$_2$-dominated clathrates in the martian soil that would have efficiently sequestered these noble gases (Mousis et al. 2010; Mousis et al. 2012). In this scenario, these noble gases would have been trapped in clathrates $\sim$4 Gyr b.p. when the CO$_2$ surface pressure was expected to be of the order of a few tens to a few hundred mbar (Mousis et al. 2012). This scenario implies that the $^{36}$Ar, $^{84}$Kr, and $^{130}$Xe abundances measured in the planet's atmosphere are not representative of its global noble gas budget. Depending on the amount of existing clathrates, the volume of noble gases trapped in these crystalline structures could be much larger than those measured in the atmosphere. In this context, two different scenarios have been proposed by Mousis et al. (2010) to explain the differences between the Ne and Ar abundances of the terrestrial planets.

In the first scenario, cometary bombardment of the planets would have occurred at epochs contemporary with the existence of their primary atmospheres. Comets would have been the carriers of Ar, Kr, and Xe, while Ne would have been gravitationally captured by the terrestrial planets (Owen et al. 1992). Only Ne and Ar would have been fractionated due to thermal and nonthermal atmospheric escape, while the abundances of the heavier noble gases would have been poorly affected by such losses. In this scenario, the combination of processes, such as escape of Ne and Ar, cometary bombardment at the epochs of existence of primary planetary atmospheres, and the sequestration of krypton and xenon in the martian clathrates, would then explain the observed noble gas abundance differences between the Earth and Mars. However, this scenario leads to an important chronological issue because depending on a the hydrogen/He amount of the captured nebula-based protoatmosphere it existed most likely only during the first few to several tens of Myr (Halliday 2003; Pepin 2006; Lammer et al. 2012).

On the other hand one should also note that heavy noble gases could have been supplied during the LHB (Marty and Meibom 2007).
In such a second scenario, Mousis et al. (2010) considered impacting comets that contained significantly smaller amounts of Ar, an idea supported by predictions of noble gas abundances in these bodies, provided that they are formed from clathrates in the solar nebula (Iro et al. 2003). Here, Ne and Ar would have been supplied to the terrestrial planets via the gravitational capture of their primary atmospheres and comets would have been the carriers of Kr and Xe only.
In this case, the cometary bombardment of the terrestrial planets could have occurred after the formation of their protoatmospheres because only the neon and argon abundances observed today would have been engendered by the escape-fractionation processes in these atmospheres.

Both scenarios preclude the possibility that material with a CI chondrite-like composition could be the main source of noble gases in terrestrial planets because the trend described by the chondritic noble gas abundances as a function of their atomic mass does not reflect those observed on Venus, the Earth, and Mars (Pepin  et al. 1992; Owen and Bar-Nun 1995). If the composition is similar to that of CI chondrites, this then excludes the hypothesis of noble gas outgassing from the interior of Mars and also the scenario of asteroidal bombardment. Irrespective of the scenario envisaged, this work does not preclude the possibility that a fraction of the heavy noble gases could have been captured by the Earth and Mars during the acquisition of nebula-based protoatmospheres. In the first scenario, the fraction of Kr and Xe accreted in this way should be low compared to the amount supplied by comets since these noble gases are not expected to have been strongly fractionated by atmospheric escape. In the second scenario, the fraction of Kr and Xe captured gravitationally by the terrestrial planets could be large if escape was efficient.

\section{Outgassing and Growth of a Secondary Atmosphere}

\subsection{Estimation of the martian water-ice reservoir by the atmospheric D/H ratio}
Volcanic outgassing is one of the main sources of volatiles for the Martian atmosphere and provides an important link between mantle and atmospheric geochemical reservoirs. Information on exchange processes between the different reservoirs is contained in the atmospheric isotopic ratios $R$ of elements such as hydrogen, carbon, and the noble gases.
$R$ may change as a function of time as lighter isotopes can escape to space more efficiently than their heavier counterparts. Overall, the efficiency of isotopic fractionation depends on the size $S$ of the considered reservoir, the total escape flux $\phi$, and the relative efficiency of isotopic escape which may be expressed by the fractionation factor $f$ (Donahue 2004).

The ratio of the sizes of the past and present reservoirs in isotopic equilibrium with the atmosphere can be calculated if the respective isotopic ratios are known and is given by Donahue (1995)
\begin{equation} \label{Eq:AM_1}
 \frac{S_{\rm t}}{S_{\rm p}}=\left(\frac{R_{\rm p}}{R_{\rm t}}\right)^{1/(1-f)}
\end{equation}
where variables with index $p$ and $t$ refer to the present and past values, respectively.
If an initial isotopic ratio is assumed for $R_{\rm t}$, the size of the reservoir when it has last been reset to this value will be obtained. This reset may have happened due to strong volcanic outgassing, delivery of additional material by impacts, or a sudden exchange with other reservoirs not in isotopic equilibrium with the atmosphere. In this way, isotopic ratios found in Martian meteorites can be used for $R_{\rm t}$ to obtain the size of the reservoir at the time of their crystallization.

In the following, we will consider the size of the water reservoir and use the isotopic ratio of deuterium (D) and atomic hydrogen (H) ($D_0/H_0$). A compilation of different D/H isotopic ratios in the martian atmosphere, terrestrial sea water as well as comets and martian meteorites is given in Table 1.
\begin{table}
D/H ratio in the martian atmosphere, the terrestrial sea water, comets and various martian
meteorites. The crystallization ages and ejection ages of the meteorites are taken
from Nyquist et al. (2001) and for ALH 84001 from Turner et al. (1997). The measured D/H values are
taken from Leshin et al. (1996) and are also given in units of terrestrial sea water D/H. The ejection age refers to the estimated time of ejection
from the Martian surface.
\begin{center}
\begin{tabular}{l|cccc}
\hline\noalign{\smallskip}
Reservoir     & D/H [$1\times 10^{-4}$] & D/H [SMOW]  &   Cryst. age [Myr]    &   Ejection age [Myr]    \\\hline
Mars atmosphere & 8.0 & 5.13 & & \\
Terrestrial sea water (SMOW) & 1.56 & & 1.00  &   \\
Comets           &  $\sim$3.2 &  $\sim$ 2.05 &  &   \\ \hline
martian meteorites  &         & &    &   \\ \hline
AH 84001     & 2.45  & 1.57 & 3920$\pm$40 & 15.0$\pm$0.8 \\
Chassigny    & 1.49-1.6 & 0.96-1.03 & 1340$\pm$50 & 11.3$\pm$0.6\\
Nakhla      & 2.24-2.73 & 1.44-1.75 & 1270$\pm$50 & 10.75$\pm$0.4\\
Lafayette   & 2.47-2.8 & 1.58-1.79 & 1320$\pm$20 & 11.9$\pm$2.2\\
Governador Valadares& 1.97 & 1.26 & 1330$\pm$10 &10.0$\pm$2.1\\
Zagami      & 3.28  & 2.10 & 177$\pm$3 & 2.92$\pm$0.15 \\
Shergotty   &3.39   & 2.17 & 165$\pm$4 & 2.73$\pm$0.2\\
Elephant Moraine 79001 & 3.86 & 2.47 & 173$\pm$3 & 0.73$\pm$0.15\\\hline
\noalign{\smallskip}
\end{tabular}
\end{center}
\end{table}
The initial D/H ratio was modified from its initial
ratio to the present one by atmospheric escape processes. One way to estimate the initial ratio on Mars is to assume that isotopic ratios on Earth and Mars were identical following accretion. Given that the amount of water present in Earth's oceans is very large, isotopic ratios have probably changed by less than 0.2 \% since accretion (Donahue 2001) and $D_0/H_0$ can be approximated by its present day value of the Standard Mean Ocean Water (SMOW). Another way to constrain $D_0/H_0$ is to calculate the Martian primordial composition from dynamical accretion models, which result in a primordial isotopic ratio $D_0/H_0$ between 1.2 and 1.6 times that of the SMOW (Lunine et al. 2003), indicating the range of uncertainty associated with this value. The current atmospheric D/H ratio on Mars was measured with a $4\;$m reflecting telescope in combination with a Fourier transform spectrometer at Kitt Peak Observatory providing a significantly fractionated value of $R_{\rm p}=5.5 \pm 2$ SMOW (Krasnopolsky et al. 1997). From high resolution spectroscopic observations of D and H Lyman-$\alpha$ emissions of the martian hydrogen corona with the Hubble Space Telescope, the fractionation factor $f$ for D and H was estimated to be $\sim$0.016--0.02 (Krasnopolsky et al. 1998; 2000). These values are significantly lower than the theoretically calculated value of 0.32 by Yung et al. (1988), resulting in larger reservoirs than previously assumed.
\begin{figure}[t]
\begin{center}
\includegraphics[width=0.75\columnwidth]{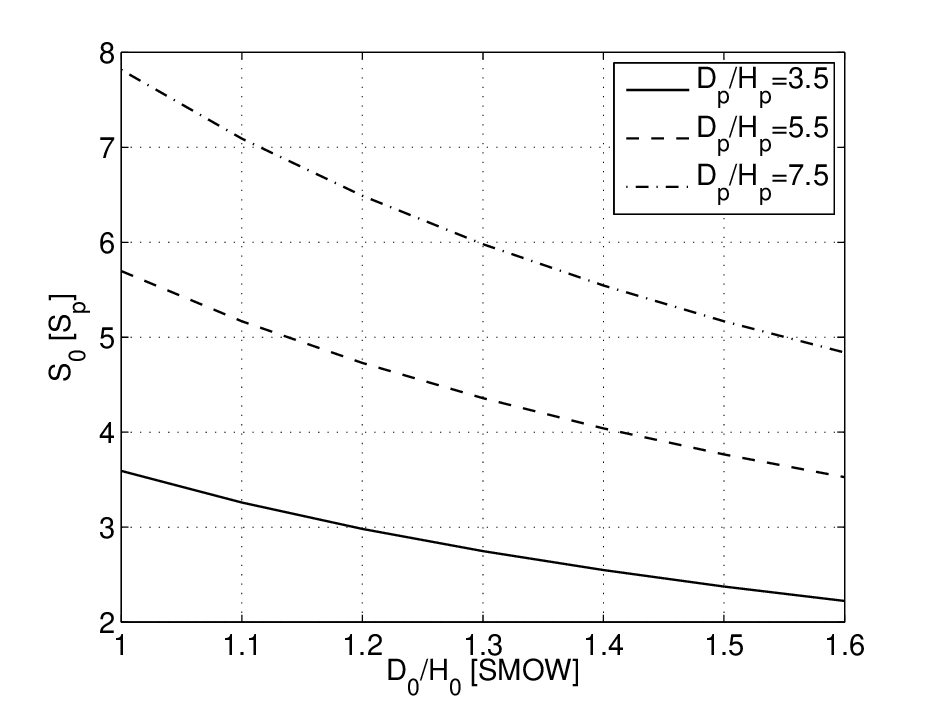}
\caption{Size of the martian water reservoir in isotopic equilibrium with the atmosphere as a function of the initial deuterium to hydrogen ratio $D_0/H_0$ for three different present-day isotopic ratios $D_p/H_p$. Reservoir size is given in terms of its present-day size $H_p$ and corresponds to the time when the isotopic ratio in the atmosphere was last reset to the initial ratio. The assumed fractionation factor $f$ is $0.02$, as measured with the Hubble Space Telescope (Krasnopolsky et al. 1998, 2000).}
\label{Fig:AM_3}
\end{center}
\end{figure}
Using $f=0.02$, the size $S_0$ of the water reservoir at the time when the isotopic ratio in the Martian atmosphere was last reset can be estimated from Eq. 2 and is shown in Fig. \ref{Fig:AM_3} as a function of the initial D/H ratio with $D_0/H_0=1$ to $1.6$. Reservoir size is given in terms of its present-day size for three different values of the current D/H ratio. This calculation implies that the past water reservoir was $2-8$ times larger than today, indicating that 50 to 88 \% of the past reservoir was lost.

In addition, the absolute size of the past water reservoir can be estimated from the total atmospheric escape flux (Donahue 2004). Results are sensitive to the assumed escape flux and early estimates arrived at a total past reservoir size corresponding to an equivalent global layer (EGL) of water between 0.2 m (Yung et al. 1988) and 30-80 m Jakosky et al. (1991). More recently, Krasnopolsky and Feldman (2001) estimated a total reservoir size of 65-120 m EGL, Lammer et al. (2003) obtained a value of $\sim$17--61 m EGL, while Donahue (2004) used isotopic ratios in the Zagami meteorite as an additional constraint to arrive at 100-800 m EGL.
A scenario where serpentinization in the crust
stored most of the ancient water reservoir has recently been proposed by Chassefi\`{e}re and Leblanc (2011c). They used the present D/H ratio to conclude that up to $\sim$400
m EGL of free water could hypothetically be stored in crustal serpentine, based on
the assumption that D and H atoms released into the atmosphere during
serpentinization have escaped and fractionated.

As the initial martian water inventory was most likely affected by hydrodynamic blow-off due to the young Sun's high EUV flux (cf. Sec. 5), H and D will have escaped unfractionated to space between $\sim$4.0--4.5 Gyr ago. Therefore, these values may give only an estimate of the amount of volatiles delivered by impacts and volcanic outgassing until $\sim$4 Gyr ago.
\subsection{Volcanic outgassing of CO$_2$ and H$_2$O}
One way to quantify the rate of volcanic outgassing is to estimate the amount of crustal production as a function of time and to multiply this volume by the magma volatile content. In this case, the outgassing rate can be obtained from
\begin{equation}
\frac{dM^{\rm atm}_{\rm i}}{dt}\propto\frac{dM_{\rm cr}}{dt}\eta X^{\rm melt}_{\rm i}
\end{equation}
where $M^{\rm atm}_{\rm i}$ is the outgassed mass of volatile species $i$, $M_{\rm cr}$ is the amount of extracted magma, $X_{\rm melt}^i$
is the concentration of volatile species $i$ in the melt, and $\eta$ is an outgassing efficiency.

As the solubility of volatiles in magmas at surface pressure is low, essentially all dissolved volatiles will be released when erupting extrusively. For intrusive volcanism, it can be assumed that plutons do not contribute to volcanic outgassing (O'Neill et al. 2007) and $R_i$ will depend on the ratio of intrusive to extrusive volcanism. However, as volatiles will be enriched in the remaining liqud during solidification, it is also possible that dissolved volatiles will outgas at depth and reach the atmosphere (Hirschmann et al. 1998). An intermediate approach is to assume that volatiles can be delivered to the surface as long as some crustal porosity is present at the depth of the intrusion (Grott et al. 2011).
\begin{figure}[t]
\begin{center}
\includegraphics[width=0.75\columnwidth]{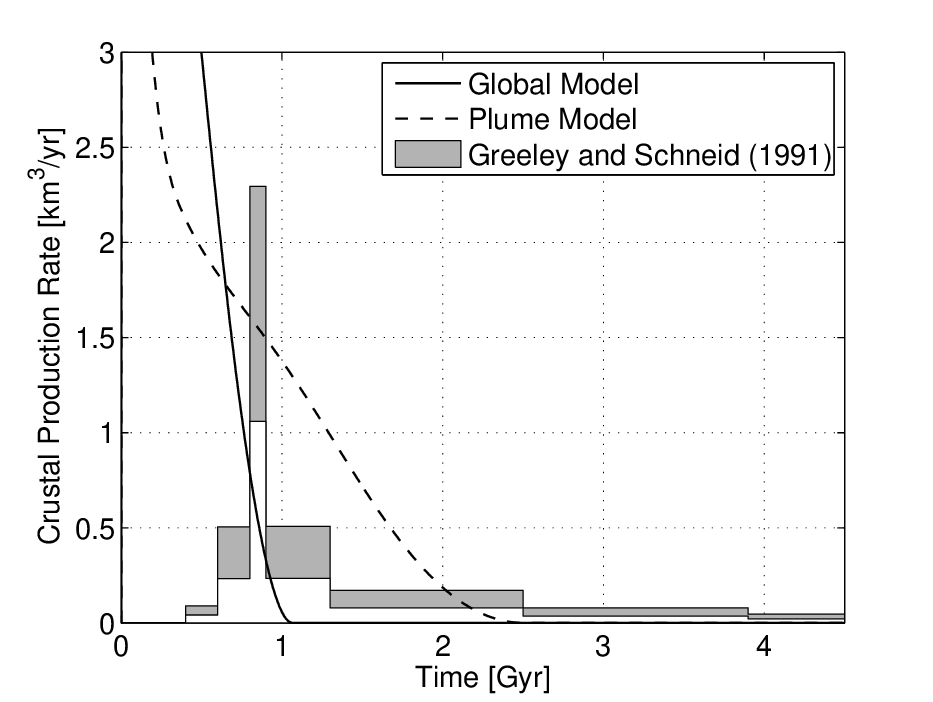}
\caption{Different estimates of crustal production rates as a function of time. The solid and dashed lines show the rates obtained with parameterized thermochemical evolution models assuming a global melt layer and melt generation in localized, hot mantle plumes, respectively (Grott et al. 2011; Morschhauser et al. 2011). The shaded area corresponds to the values obtained from photogeological estimates by Greeley and Schneid (1991) with ratios of extrusive to intrusive volcanism ranging from 1:5--1:12.}
\label{Fig:AM_4}
\end{center}
\end{figure}
The volume of crustal production has been estimated from the photogeological record and crater counting has been used to age-date the corresponding surfaces. In this way, the rate of crustal production, i.e., the amount of crust produced as a function of time, has been determined. Lava volumes were obtained from the topology of partially filled impact craters by comparing their actual depth to the theoretical values obtained from depth-to-diameter scaling relations (Greeley and Schneid 1991). As only extrusions can be assessed, estimates of the ratio of extrusive to intrusive volcanism are necessary to obtain the total volume of produced crust. Also, older deposits may be covered by later extrusions and therefore early crust production rates may be underestimated.

Alternatively, numerical models of the thermochemical evolution of Mars can be used to calculate the globally averaged crustal production rates (Hauck and Phillips 2002; Breuer and Spohn 2006; Fraeman and Korenaga 2010; Morschhauser et al. 2011). Note, however, that parameterized models cannot account for lateral variations in crustal production, and fully two- or three dimensional models need to be applied in order to resolve young, localized volcanism. Fig. \ref{Fig:AM_4} compares the photogeological estimates of Greeley and Schneid (1991) with the results of the numerical models by Grott et al. (2011) and Morschhauser et al. (2011), who assume melt production in a global melt layer and melt production in localized, hot mantle plumes, respectively. Within uncertainties of the ratio of extrusive to intrusive volcanism, these approaches show satisfactory agreement at intermediate epochs, but the photogeological approach underestimates crustal production rates in the Noachian, whereas the numerical models cannot provide estimates for younger volcanism.
Therefore, both approaches complement each other and are necessary for an overall picture of Mars'
volcanic history.

As Mars is in the stagnant-lid mode of mantle convection, volatile contents of magmas associated with intra-plate volcanism on Earth have been considered to be comparable to volatile contents on Mars. However, as Mars may have a different
volatile content and mantle oxidation state compared to Earth, these
values ar at best first-order estimates. However, for lack of better data at the time, terrestrial values fave been assumed in several studies. Today, a better understanding of the
differences of the magma volatile content of Earth and Mars exist and a
more sophisticated approach will be elaborated at the end of this section.
For Mauna Loa and Kilauea on Hawai'i, the concentration of CO$_2$ and H$_2$O in the magma is $0.65\;\textrm{wt.}\%$ and $0.30\;\textrm{wt.}\%$, respectively (Gerlach and Gr\"{a}ber 1985; Greenland 1987a; 1987b). Subglacial volcanism on Iceland with a CO$_2$ content of $\sim500$ ppm has also been used as an analogue for Mars (O'Neill et al. 2007), although it may differ significantly from the type of volcanism expected in the stagnant-lid regime of mantle convection.
Using Hawai'ian volcanism as an analogue for Mars and by neglecting atmospheric escape, Phillips et al. (2001) concluded that 1.5 bar of CO$_2$ have been outgassed during the formation of the Tharsis bulge including intrusions, probably leading to a strong greenhouse effect and climate transition at the end of the Noachian. Since the mid-Noachian, extrusive volcanism may have outgassed an atmosphere of 800 mbar, consisting of 400 mbar CO$_2$, 8 m EGL H$_2$O and 6 other minor species as has been estimated from the photogeological record and Hawai'ian volatile concentrations (Craddock and Greeley 2009).

Being of igneous origin, the volatile content of the Martian meteorites may also serve as a proxy for magma volatile contents. An analysis of melt inclusions and a reconstruction of SNC solidifcation history results in a magma water content of $1.4-1.8 \;\textrm{wt.}\%$ prior to degassing (McSween and Harvey 1993; McSween et al. 2001; Johnson et al. 1991). As an upper bound, the formation of Tharsis could have outgassed 120 m EGL H$_2$O in this way, if a magma volatile content of 2 wt.\% H$_2$O is assumed (Phillips et al. 2001). However, water content may have been overestimated and values change to less than $0.3\;\textrm{wt.}\%$ if the high chlorine content in Martian meteorites is taken into account (Filiberto and Treiman 2009). These lower water concentrations are also supported by direct measurements in kaersutitic and biotitic melt inclusions (Watson et al. 1994) and would decrease the amount of water outgassed by Tharsis to 18 m EGL.

It may be argued that the magma water content of the SNC meteorites,
which are believed to be younger than $\sim$1.3 Gyr, do not represent the typical
magma water content at the time of Tharsis formation $\sim$3 Gyr ago.
As numerical models predict a total mantle water loss of $\sim$50\% due to
volcanic outgassing, the uncertainty associated in using SNC magma water
contents for Tharsis outgassing is around a factor of two. Compared with
other uncertainties, e.g. the volume of Tharsis or the debate on the water
magma concentration of the SNCs, this uncertainty is not significant.

In addition, partitioning of water into the melt can be calculated from a melting model. Accumulated fractional melting may be more appropriate for mantle melting on Mars (Fraeman and Treiman 2010; Grott et al. 2012, this issue), but batch melting is also widely applied (Hauck et al. 2002; Morschhauser et al. 2011). However, the difference between the two approaches is comparatively small and results are not significantly affected. Within the frameworks of batch- and fractional melting, the partioning coefficient, melt fraction, and bulk water content determine the concentration in the melt. The partition coefficient of water is most likely close to $0.01$ (Katz et al. 2003), and melt fractions obtained from numerical models average around $5-10\;\%$ (Hauck et al. 2002; Morschhauser et al. 2011). These values are consistent with melt fractions determined from trace-element analysis of shergottites, which range from $2\;\%$ to $10\;\%$ (Norman 1999; Borg and Draper 2003) and result in magma water concentrations of 10 to 16 times the bulk water content. The bulk mantle water content of Mars is poorly constrained, and a large range of concentrations have been obtained from different methods: Analysing water content in melt inclusions of Martian meteorites, a bulk mantle water concentration of $1400\;\textrm{ppm}$ was calculated (McSween and Harvey 1993), while numerical accretion models (Lunine et al. 2003) arrive at maxiumum concentrations of $800\;\textrm{ppm}$. In contrast, meteoritic mixing models constrained from element ratios in SNC meteorites predict bulk water concentrations of only $36\;\textrm{ppm}$ (W\"{a}nke and Dreibus 1994).
\begin{figure}[t]
\begin{center}
\includegraphics[width=0.95\columnwidth]{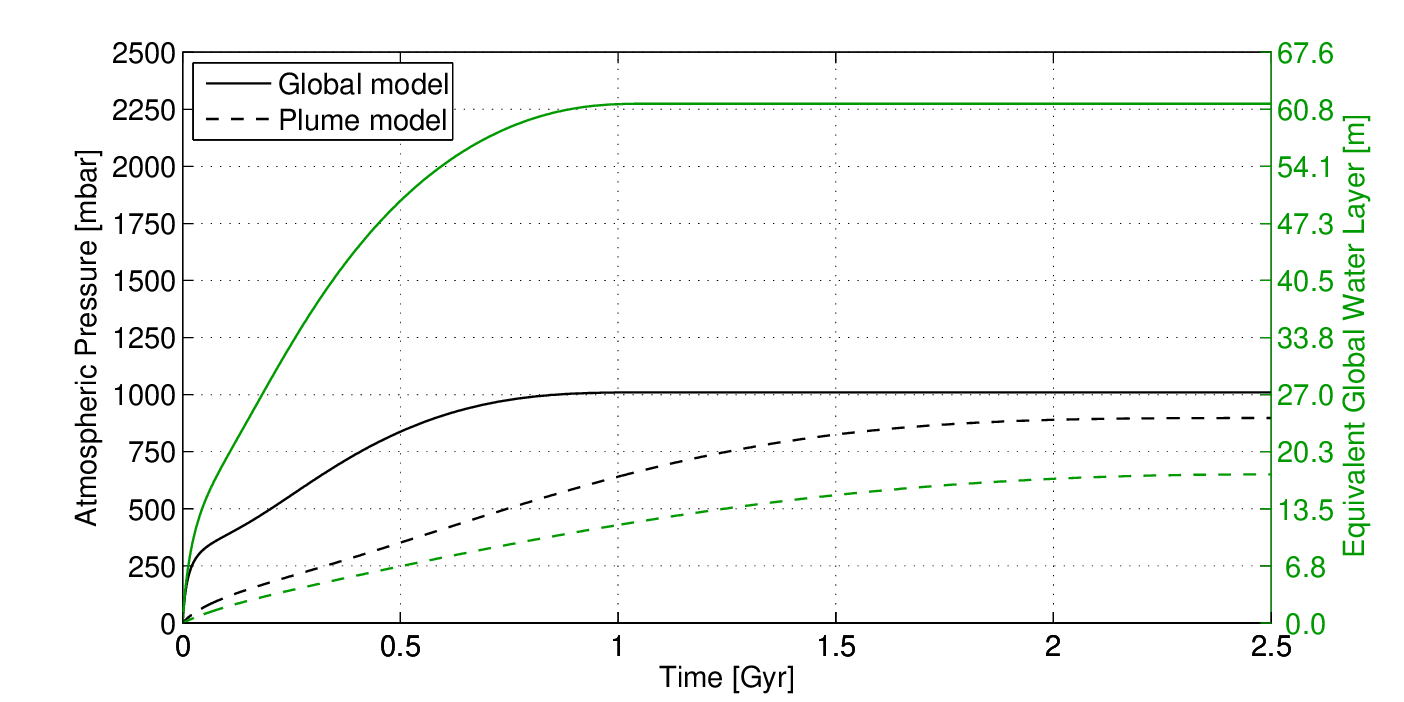}
\caption{Modeled cumulative volcanic outgassing of CO$_2$ given as partial surface pressure in mbar (black lines) and of H$_2$O given as equivalent global water layer
(green lines) as a function of time. Mantle oxygen fugacity was assumed to be one order of magnitude above the iron-wustite buffer, resulting in an upper limit on CO$_2$ pressure. Initial mantle water concentration was assumed to be $100\;\textrm{ppm}$. Solid curves correspond to a model considering mantle melting in a global melt channel, whereas dashed curves correspond to a model considering melting in mantle plumes covering only a small fraction of the planetary surface. The outgassing efficiency $\eta$ was set to $0.4$ for all models
and atmospheric escape is neglected.}
\label{Fig:AM_5}
\end{center}
\end{figure}
It should be noted that, even though the mantle
is dehydrating with time and SNCs are believed to be geologically young,
the inferred bulk mantle water content of the SNCs is larger than that predicted
by the other methods. This may be due to the large uncertainties
associated with each of these methods.

The solubility of CO$_2$ in Martian magma can be calculated by considering the underlying chemistry. For CO$_2$, solubility depends on the form in which graphite is stable in the Martian mantle, which in turn depends on oxygen fugacity (Hirschmann and Withers 1998). The Shergottites, which most likely reflect conditions at the magma source region (Hirschmann and Withers 1998), have oxygen fugacities between the iron-wustite (IW) buffer and one $\log_{10}$ unit above it (IW+1) (Herd et al. 2002; Shearer et al. 2006). The oldest Martian meteorite, ALH84001, indicates even more reducing conditions around IW-1 (Warren and Kallmeyen 1996). Under these reducing conditions, carbon is stable in the form of graphite (Hirschmann and Withers 1998), and a chemical model for CO$_2$ solubility under graphite saturated conditions (Holloway et al. 1992; Holloway 1998) can be applied (Hirschmann and Withers 1998). Chemical equilibrium constants controlling CO$_2$ solubility have been calibrated using terrestrial basaltic magmas (Holloway et al. 1992) and Martian-basalt analogue material (Stanley et al. 2011), as a function of oxygen fugacity. At relatively oxidizing conditions (IW+1) and melt fractions typically encountered in the Martian mantle (5-10 \%), a maximum of $\sim1000\;\textrm{ppm}$ CO$_2$ can be dissolved (Grott et al. 2011; Hirschmann and Withers 1998), which is significantly less than the $0.65\;\textrm{wt.}\%$ obtained for Kilauea basalts.

Combining the chemical model for CO$_2$ solubility (Hirschmann and Withers 1998) with parameterized thermal evolution models (Morschhauser et al. 2011), the amount of outgassed CO$_2$ can be calculated self-consistently (Grott et al. 2011). In order to cover the range of expected mantle dynamics in a one-dimensional model, two end member melting models may be considered: Melting in a global melt channel is likely representative for early martian evolution, whereas melting in localized mantle plumes may be more appropriate for the later evolution. Outgassing for both models is shown in Fig. \ref{Fig:AM_5}. The degree of partial melting encountered in the plume model is generally higher than that for the melt channel model, and volatile concentrations in the melt are therefore lower (cf. Fig. 5). As approximately the same amount of crust is extracted from the mantle in both models, this results in reduced outgassing efficiencies. In both cases, a total of $\sim$1 bar CO$_2$ can be outgassed if comparatively oxidizing conditions (IW+1) are assumed, and for an initial mantle water concentration of $100\;\textrm{ppm}$ a total of 61 and 18 m EGL of H$_2$O can be outgassed in the global melt channel and plume model, respectively. While the rate of outgassing is lower for the plume model, outgassing in this model persists for $\sim$1 Gyr after melt generation ceases in the global melt channel model. A parameterization of CO$_2$ and H$_2$O outgassing rates as a function of oxygen fugacity, initial mantle water content, and outgassing efficiency may be found in Grott et al. (2011).

From these considerations one can see that, depending on geochemical and geological constraints,
early Mars could have accumulated a secondary CO$_2$ atmosphere by volcanic outgassing of $\leq$1 bar $\sim$4 Gyr ago. However, large impacts and
atmospheric escape processes should have modified the growth of this secondary atmosphere. In the following sections we will investigate
possible changes of the secondary atmosphere in relation to losses and sources caused by large impacts, as well as
various atmospheric escape processes which are connected to the change in solar activity.

\section{Atmosphere Erosion and Delivery by Large Impacts}
Atmospheric erosion and delivery by impactors can  be studied with the help of hydrocode simulations which essentially simulate the flow field and dynamic response of
materials by taking into account material strength and rheology (Pierazzo and Collins 2003; Shuvalov and Artemieva 2001; Svetsov 2007, Melosh and Vickery 1989). Previous
atmospheric erosion studies by hydrocodes have not always provided similar results, mainly due to differences in the physical models such as the choice of an appropriate
equation of state, or a proper model of vapor cloud dynamics (Pham et al. 2009). In addition, these simulations require very large computer resources and can not be used
directly to simulate long term atmospheric evolution. Therefore, the influence of the major parameters on atmospheric erosion and delivery has been parameterized. Models using the parameterization of the major mechanisms affecting the atmospheric erosion and delivery  by the impacts  can be instead used to study the evolution
of the atmosphere.  Many studies applied the so-called ``tangent plane model'' which has been developed by Melosh and Vickery (1989).

The tangent plane model of Melosh and Vickery (1989)  is based on their hydrocode simulation results. Their model has been modified to take into account other simulations as well as additional parameters, and has been used to  obtain a global view of the atmospheric mass evolution (Zahnle et al. 1992; Zahnle 1993; Manning et al. 2006; 2009; Pham et al. 2009).
The advantage of using analytical models is that they can represent basic aspects of impact erosion and delivery, while reducing computation time since
they only use a reduced number of parameters, scaled with numerical hydrocode simulation results.

The principle of the tangent plane model is that, when an impactor above a critical mass, $m_{\rm crit}$, strikes the planet,
the total mass above the plane tangent to the surface at the impact point, $m_{\rm tan}$, escapes.

The critical mass is the minimal
impactor mass that can eject $m_{\rm tan}$, and it is proportional to $m_{\rm tan}$ through a factor $n$ which represents the impact efficiency.
The atmospheric mass above the plane tangent of the impact surface is approximated by $m_{\rm tan} =  m_{\rm atm} H /2R_{\rm pl}$, assuming
an isothermal atmosphere in hydrostatic equilibrium, where $m_{\rm atm}$  is the total atmospheric mass, $H$ the atmospheric scale height,
and $R_{\rm pl}$ the radius of the planet. Note that the tangent plane model is only an approximation of erosion and delivery processes
and that small impactors can still remove atmosphere (Zahnle, 1993). While the model doesn't reproduce the physics of
impact erosion or delivery, it can, with a suitable parameterization of the critical mass, give a global view
of the atmospheric mass evolution upon impacts with a minimum set of variables (related to the critical mass value) and a much smaller computation time.
\begin{table}
 \centering
\begin{tabular}{c|c|c|c}
\hline
Factor & Asteroids & \multicolumn{2}{c}{Comets} \\
& & SP comets & LP comets \\
\hline
$f_{vel}$ & 0.08 & 0.83 & 0.99 \\
$f_{obl}$ & 7.54 & \multicolumn{2}{c}{2.16} \\
$y_{imp}$ & 0.01 & \multicolumn{2}{c}{0.3} \\
$f_{vap}$ & 0.34 & \multicolumn{2}{c}{1} \\
$g_{vap}$ & 0.21 & \multicolumn{2}{c}{1} \\
\hline
\end{tabular}
\caption{Values of the factors given in Eqs. (6) and (7).}\label{coeffimpacts}
\end{table}
\begin{figure}[h]
\begin{center}
\includegraphics[width=0.8\columnwidth]{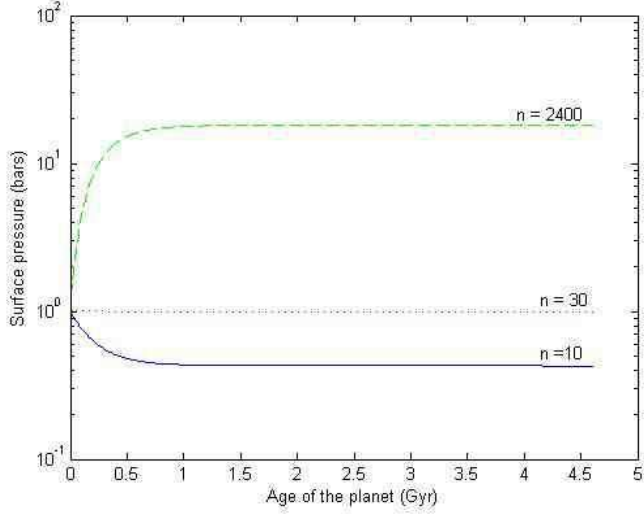}
\caption{Maximum diameter of the impactors hitting Mars as a function of time, for an assumed initial atmospheric
surface pressure of 300 mbar. The horizontal lines show the critical diameter $d_{\rm crit}$ for different values of
$n$  = $m_{\rm crit}/m_{\rm tan}$ (details in the main text). Only impactors with diameter larger than $d_{\rm crit}$ can erode
the atmosphere. The lower limit of $n=10$ represents an extreme case corresponding to the  upper limit of atmospheric erosion.}
\end{center}
\end{figure}
The total mass evolution $M_{\rm atm}$ is controlled by the difference between the  rates of change of atmospheric erosion, $M_{\rm esc}$
and the delivered volatile mass, $M_{\rm del}$
\begin{equation}
\frac{dM_{\rm atm}}{dt}=\frac{dM_{\rm del}}{dt}-\frac{dM_{\rm esc}}{dt},
\end{equation}

with

\begin{equation} \label{Mesc2}
\frac{dM_{\rm esc}}{dt}=\frac{\partial N_{\rm cum}[>m_{\rm crit}(t),t]} {\partial t} 4 \pi R^2 m_{\rm tan}(t)f_{\rm vel}f_{\rm obl},
\end{equation}

and

\begin{eqnarray} \label{Mdel2}
\frac{dM_{\rm del}}{dt} =\frac{\partial N_{\rm cum}[>m_{\rm crit}(t),t]} {\partial t}  4 \pi R^2 \frac{b}{1-b} m_{crit}(t) y_{imp} f_{\rm vap} + \nonumber \\
 \frac{\partial N_{\rm cum}[>m_{\rm crit}(t),t]} {\partial t}  4 \pi R^2 m_{crit}(t) y_{imp} (1-f_{\rm vel} f_{\rm obl}  g_{\rm vap}).
 \end{eqnarray}
where $N_{\rm cum}$ is the cumulative number of impacts with mass larger than $m_{\rm crit}$ at a time $t$, $\partial N_{\rm cum}(>m_{\rm crit}(t),t) / \partial t$
represents the flux of these impactors and $b$ characterizes the mass distribution of the impactor flux, $b<1$. We also assumed an exponentially decaying impact flux (e.g., Neukum and Wise 1976; Ivanov 2001; Neukum et al. 2001).

In the above equations the original ``tangent plane model''  is modified by the additional terms  $f_{\rm vel}$ and $f_{\rm obl}$ in Eq. \ref{Mesc2} and $y_{\rm imp}$  $f_{\rm vap}$, $g_{\rm vap}$, $f_{\rm vel}$ and $f_{\rm obl}$ in Eq. \ref{Mdel2} which were not considered in previous studies (Pham et al. 2009; 2011). The fraction of impactors which are fast enough to erode the planet ($f_{\rm vel}$) as well as the enhancement factor of the erosion due to impact obliquity ($f_{\rm obl}$)  are taken account.  The volatile content ($y_{\rm imp}$) is different for asteroids and comets. We assume a volatile content of $y_{\rm imp}$=0.03 for comets, and $y_{\rm imp}$=0.01 wt. for asteroids. The ratio of the vaporized mass to the impactor mass averaged over impact velocities is considered through the parameters $f_{\rm vap}$ or $g_{\rm vap}$ depending on whether the impactor mass is below or above the critical mass, respectively. In the delivery equation, the first term on the right side represents the delivered mass rate for $m_{\rm imp}< m_{\rm crit}$, and  the second term is the delivered mass rate for the fraction of impactors with $m_{\rm imp} > m_{\rm crit}$  that was not removed by impacts. The relative amount of comets and asteroids in the total impact  flux on Mars is assumed to be  6\% and 94\%, respectively, as suggested by  Olsson-Steel (1987). In addition comets are differentiated between short-period (SP) comets ($\sim$4\%) and long-period comets (LP) from the Oort cloud ($\sim$2\%) when the impact velocity is taken into account in the simulations.   The values corresponding to the factors given in Eq. \ref{Mesc2} and Eq. \ref{Mdel2} are shown in Table 2.

The efficiency of atmospheric erosion and delivery is determined by the factor $n$ which is given by $n= m_{\rm crit}/m_{\rm tan}$. The exact value of $n$ in
the tangent plane model is uncertain. The value suggested initially by Melosh and Vickery (1989), $n=1$, has been revised by more recent studies to $n=10$ (Vickery 1990; Manning
et al. 2006). On the other hand, the more recent hydrocode simulations performed by Shuvalov and Artemieva (2001),
Ivanov et al. (2002), and Svetsov (2007) yield results compatible with much larger values of the critical mass, on the order of $50< n<2000$.
Note that $n$ is a function of atmospheric pressure (Hamano and Abe 2006; Svetsov 2007). Pham et al. (2009) considered the tangent plane model with different impact erosion efficiencies, using constant as well as pressure-dependent values of $n$.

The total atmospheric mass evolution calculated by using Eqs. \ref{Mesc2} and \ref{Mdel2}
is plotted in Fig. 5 for three values of $n$. Depending on the value of  $n$  impacts constitute either a factor of  erosion or a source of volatiles on Mars. The most recent hydrocode simulations tend to favor larger values,  $n$ $>$ 30, for which impacts are a source of volatile (Svetsov, 2007). Although impacts can remove atmosphere for smaller values of $n$ the process is not very efficient, since even for the most favorable values of $n=10$, a 1 bar protoatmosphere can not be eroded to present values over the age of the Solar System.  In Sect. 6  we will consider the most favorable parameters for atmospheric loss  considering a large impact flux during the late heavy bombardment with lower limit of $n$ to yield an upper limit of atmospheric erosion.  Although  the impact erosion may not have been a very relevant loss process the question remains if the impact flux for accumulation of a secondary atmosphere was higher than the expected EUV-powered thermal escape flux which will be discussed in Sect. 5.

\section{The EUV-Powered Blow-Off of the Protoatmosphere and the Change of a Secondary CO$_2$ Atmosphere by Escape Processes}
From observations of young solar-type G stars it is known that despite a weaker total luminosity,
stars with a young age are a much stronger source of X-rays and EUV electromagnetic
radiation (e.g., Newkirk 1980; Skumanich and Eddy 1981; Zahnle and Walker 1982; G\"{u}del et al. 1997).
Since the 90ies the evolution of UV fluxes of a sample of solar analogue stars, so-called proxies of the Sun
have been studied in detail by spectral measurements from the IUE satellite (Dorren and Guinan 1994).
This research was extended by G\"{u}del et al. (1997), Ribas et al. (2005) and recently by Claire et al. 2012
to X-rays and EUV.
The wavelength range $\lambda\leq$1000{\AA} is relevant for ionization, dissociation
and thermospheric heating (e.g., Hunten 1972; 1987; 1993).

Because X-rays dominate at young stellar ages, Owen and Jackson (2012) studied the contribution of harder X-ray's to
the heating of hydrogen-rich upper atmospheres. If we compare the X-ray luminosities of solar proxies with younger age
(e.g. Ribas et al. 2005; Claire et al. 2012) at Mars' orbit, with the values necessary for having a dominating X-ray driven
atmospheric escape (Owen and Jackson 2012, their Fig. 11), one finds that this process is only relevant for hydrogen-rich
``Hot Jupiter''-type exoplanets but can be neglected on early Mars which orbits further away from the Sun. Therefore,
for early Mars, EUV radiation should be the main heating process in the thermosphere.

G\"{u}del et al. (1997),
Ribas et al. (2005) and Claire et al. 2012 analyzed multi-wavelength EUV observations
by the ASCA, ROSAT, EUVE, FUSE and IUE satellites of solar proxies with ages $<$4.6 Gyr and found that the EUV flux is saturated
during the first 100 Myr at a value $\sim$100 times that of the present Sun. As shown in Fig. 7
this early active period of the young Sun decreases according to an EUV enhancement factor power law
after the first 100 Myr (Ribas et al. 2005)
\begin{equation}
S_{\rm EUV}=\left(\frac{t_{\rm Gyr}}{t_{0}}\right)^{-1.23},
\end{equation}
where $t_{0}$ is the age of the present Sun and $t_{\rm Gyr}$ the younger or older age of the
Sun in time in units of Gyr.

Due to the lack of accurate astrophysical observations from solar proxies with different ages,
previous pioneering studies on EUV-driven hydrodynamic escape of primitive atmospheres were based
only on rough EUV enhancement scaling factors which were assumed to be similar or up to only
$\sim$5--25  times higher than the present value (e.g., Watson et al. 1981; Kasting and Pollack 1983;
Chassefi\`{e}re 1996). Furthermore, most of the previous studies are based on terrestrial
planet formation models where the accretion for Mars occurred quite late at $\sim$100 Myr
(Wetherill 1986).

Because Mars can be considered as a planetary embryo that did not collide or merge with other embryos
it developed most likely within $\sim$2--4 Myr after the birth of the Solar System (Dauphas and Pourmand 2011; Brasser 2012).
This age agrees with average planetary nebula evaporation time scales of $\sim$3 Myr. Because nebulae
life times are $<$10 Myr (e.g., Lunine et al. 2011), the nebula-based hydrogen-rich martian protoatmosphere was most likely exposed during several tens of Myr or
up to $\sim$150 Myr to an EUV flux which was $\sim$50--100 times higher compared to today's Sun (Ribas et al. 2005; Claire et al. 2012).
\begin{figure}[h]
\begin{center}
\includegraphics[width=0.75\columnwidth]{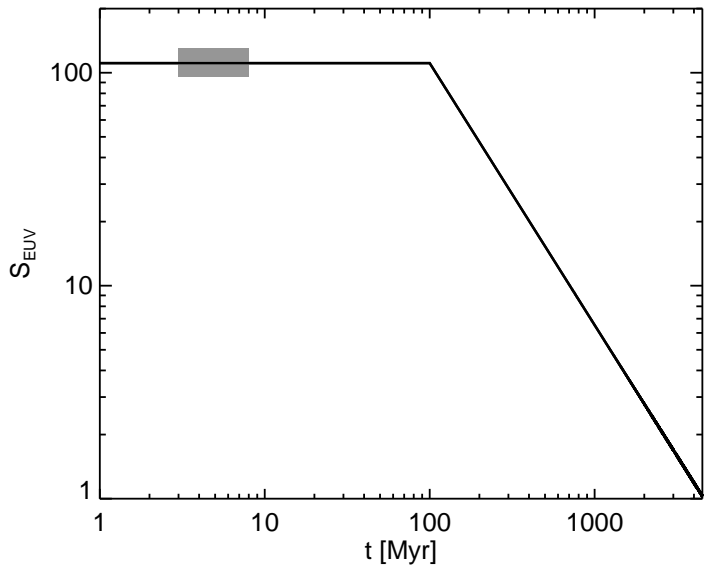}
\caption{Solar EUV flux enhancement factor $S_{\rm EUV}$ as obtained from observations of solar proxies.
The average nebula evaporation time is $\sim$3 Myr. Since that time period planetary embryos and protoplanets
are exposed to the saturated EUV flux value which is $\sim$100 times larger compared to the present solar value for about 90 Myr.
The shaded area marks the expected time when Mars ended its accretion.}
\end{center}
\end{figure}
Because of the high EUV flux of the young Sun H$_2$, H$_2$O and most CO$_2$ molecules in the thermosphere are dissociated
and H atoms should dominate the upper atmosphere until they escaped to space.

Depending on the composition of the upper atmosphere and the planet's mean density,
when the solar EUV flux in the wavelength range $\lambda\approx$2--120 nm overcomes a critical value,
the outward flow of the bulk thermosphere cools due to adiabatic expansion (Tian et al. 2005; 2008).
According to studies of Watson et al. (1981), Kasting and Pollack (1983) and
Tian et al. (2005), if hydrogen populates the upper atmosphere of a terrestrial planet, its
exobase level can expand several planetary radii if the EUV flux is only a few times higher
compared to that of today's Sun. A hydrogen-rich upper atmosphere of a martian-type body which is
exposed to a EUV flux which is 5--100 times higher compared to the present solar value is,
therefore, certainly in the blow-off regime. Under such conditions the exosphere evaporates
as long as enough hydrogen is present.

By applying a blow-off formula which is derived from the energy-limited equation (e.g., Hunten 1987; 1993)
the atmospheric mass loss $dM_{\rm esc}/dt$ can be written as
\begin{equation}
\frac{dM_{\rm esc}}{dt}=\frac{3\eta S_{\rm EUV}F_{\rm EUV}}{4G\rho_{\rm pl}},
\end{equation}
 where the heating efficiency $\eta$ is the ratio of the net heating rate to the rate of solar EUV
energy absorption of $\sim$15--40\% (Chassef\`{\i}ere 1996; Lammer et al. 2009; Koskinen et al. 2012), the gravitational constant $G$, the mean planet
density $\rho_{\rm pl}$ and the present time EUV flux $F_{\rm EUV}$ in Mars' orbit. Using this relation one can estimate the atmospheric
escape as long as hydrodynamic blow-off conditions occur which means that the thermal energy of the gas kinetic motion
overcomes the gravitational energy.

Fig. 8a shows the upper limit of a EUV-driven hydrogen-dominated protoatmosphere, which may have been captured from the nebula,
during $\sim$1 Gyr after the planet's origin. The loss is estimated from Eq. (8) with a heating efficiency $\eta$ of 40 \% (Koskinen et al. 2012).
One can see that Mars could have lost an equivalent hydrogen content as available in $\sim$14 Earth oceans (EO$_{\rm H}$).
Due to the low gravity of Mars and EUV fluxes on the order of $\geq$50 times that of the
present Sun, the blow-off condition was 100 \% fulfilled for light hydrogen atoms and most likely also for heavier atomic species such as
O, or C if they populated the upper atmosphere (e.g., Tian et al. 2009).

After the dissociation of the H$_2$O molecules and a fraction of CO$_2$ one can expect that O atoms are the major form of escaping
oxygen. The escape flux of the heavier atoms $F_{\rm heavy}$ which can be dragged by the dynamically outward flowing hydrogen
atoms with flux $F_{\rm H}$ can be written as (e.g., Hunten et al. 1987; Chassefi\`{e}re 1996)
\begin{equation}
F_{\rm heavy}=\frac{X_{\rm heavy}}{X_{\rm H}}F_{\rm H}\left(\frac{m_{\rm c}-m_{\rm heavy}}{m_{\rm c}-m_{\rm H}}\right),
\end{equation}
where $X_{\rm H}$ and $X_{\rm heavy}$ are the mole mixing ratios. $m_{\rm H}$ and $m_{\rm heavy}$
are the masses of the hydrogen atom and the heavy species. $m_{\rm c}$ is the so called cross over mass
\begin{equation}
m_{\rm c}=m_{\rm H}+(kTF_{\rm H})(bgX_{\rm H}),
\end{equation}
which depends on $F_{\rm H}$, a molecular diffusion parameter
$b$ (Zahnle and Kasting 1986; Chassefi\'{e}re, 1996), the gravity acceleration $g$, Boltzmann constant
$k$ and an average upper atmosphere temperature $T$, which can be assumed under such conditions for hydrogen to be on the order of $\sim 500$ K (Zahnle and Kasting 1986; Chassefi\'{e}re 1996).
\begin{figure}[h]
\begin{center}
\includegraphics[width=0.49\columnwidth]{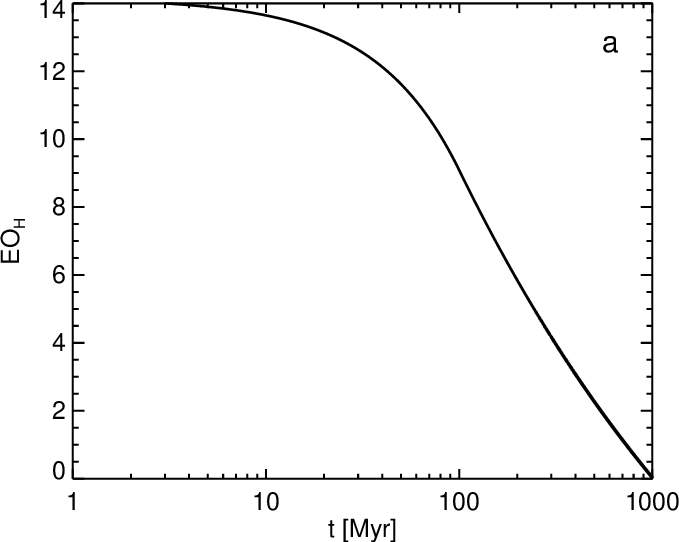}
\includegraphics[width=0.49\columnwidth]{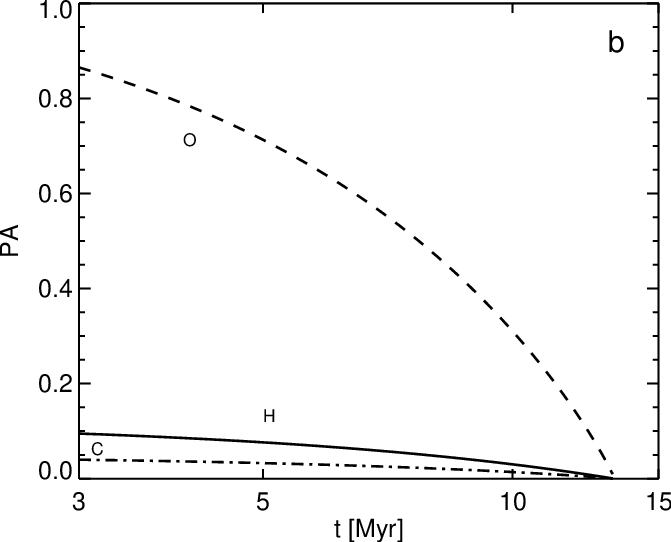}
\caption{a: Upper escape value of atomic hydrogen in units of Earth ocean equivalent amounts (EO$_{\rm H}$) of atomic hydrogen from a martian protoatmosphere
between 3 Myr to 1 Gyr after the origin of the Solar System.
b: Calculated normalized loss of an outgassed 70 bar water vapor and 12 bar CO$_2$ steam atmosphere as a function of time. PA=1 corresponds to the total pressure of 82 bar, while the solid line corresponds to loss of the hydrogen content. The dashed and dash-dotted lines corresponds to dragged oxygen and carbon atoms which originate from dissociation of H$_2$O and CO$_2$.}
\end{center}
\end{figure}
By applying Eqs. 1 and 2 and assuming that Mars finished its accretion within the EUV-saturated epoch of
the young Sun, Fig. 8b shows the atmosphere loss estimation of an outgassed water vapour dominated by a 70 bar H$_2$O and
12 bar CO$_2$ steam atmosphere. The loss is normalized to the total outgassed surface pressure of 82 bar, where PA=1 which corresponds to the total pressure of 82 bar. One can see that under such conditions early Mars could easily lose its initial atmosphere in $\sim$10 Myr. Thus, because Mars accreted early (Dauphas and Pourmand 2011; Brasser 2012),
even if the planet would have obtained its volatile inventory later, the high EUV flux of the young Sun would have blown the atmosphere away.
One can also see that under these extreme conditions the outgassed CO$_2$ would be
lost, either in dissociated form as C and O as shown in Fig. 8b. Therefore, a dense CO$_2$ atmosphere could have been
lost very early and could not have accumulated during the early Noachian. On the other hand, if early Mars was
surrounded by a nebula-based hydrogen envelope, the outgassed heavier volatiles may have been protected against atmospheric
escape until the captured hydrogen was lost and did not dominate the upper atmosphere anymore.

For these time scales the outgassed H$_2$O/CO$_2$ atmosphere remained most likely in steam form because
the time scale where the surface temperatures may reach the point that H$_2$O can condense is comparable
(Elkins-Tanton 2008). Furthermore, these time scales also agree with
studies by several researchers who investigated the early stages of accretion and impacts and expect that due
to thermal blanketing hot temperatures could keep the volatiles in vapor phase for several tens of Myr or even up to
$\sim$100 Myr (e.g., Hayashi et al. 1979; Mizuno et al. 1980; Matsui and Abe, 1986;
Zahnle et al. 1988; Abe 1997; Albar\`{e}de and Blichert-Toft 2007). One can also see from Fig. 5 and Fig. 6 and the discussions in Sect. 4 that during this early evolutionary period most of the impacts occurred.
Although the loss effect of these impacts may have been not so efficient when Mars had a dense outgassed H$_2$O/CO$_2$
steam atmosphere. As mentioned in Sect. 4 impacts have contributed to a permanent heating of the atmosphere.
On the other hand volatiles which were brought in to the atmosphere by large
impacts should also have been lost due to the strong hydrogen escape.

These results are in agreement with the non-detection of carbonates by the OMEGA instrument on board of ESA's Mars
Express spacecraft (Bibring et al. 2005).
Mars Express mapped a variety of units based on areas exhibiting hydrated minerals, layered deposits, fluvial floors, and ejecta of deep
craters within Vastitas Borealis with a surface
resolution in the $\sim$1--3-km range. Besides CO$_2$-ice in the
perennial southern polar cap, no carbonates were reported. Bibring et al. (2005) concluded that the non-detection of carbonates would indicate that no major
surface sink of CO$_2$ is present and the initial CO$_2$, if it represented a much higher content,
would then have been lost from Mars early rather than
stored in surface reservoirs after having been dissolved in long-standing bodies of water.

However, so far it is not clear when
the outgassing flux from Mars interior exceeded the expected escape flux so that a secondary
CO$_2$ atmosphere could grow during the later Noachian.
It should be noted that the accumulation of both the secondary outgassed atmosphere and
volatiles which were possibly delivered by later impacts is highly dependent on
atmospheric escape after the strong early hydrodynamic loss during the EUV-saturation phase
of the young Sun.

Tian et al. (2009) applied a 1D multi-component hydrodynamic thermosphere-ionosphere
model and a coupled electron transport-energy deposition model to Mars and found that for EUV fluxes
$>$10 times that of today's Sun, CO$_2$ molecules dissociate efficiently resulting in less
IR-cooling of CO$_2$ molecules in the thermosphere so that a CO$_2$ atmosphere
was most likely also not stable on early Mars after the EUV-saturation phase ended.
According to this study, the flux of the produced C and O atoms is $>$$10^{11}$ cm s$^{-1}$
before $\sim$4 Gyr ago and was of the same order as the fluxes from volcanic outgassing
(Tian et al. 2009).

Although this result seems logical, because the results of Tian et al. (2009) are model dependent
and contain various uncertainties we estimate the possible growth of such a
secondary CO$_2$ atmosphere from the
outgassing rates shown in Fig. 5  or Grott et al. (2011), by considering that the outgassed CO$_2$ flux
exceeded the thermal escape since 4.3, 4.2, 4.1 or 4 Gyr ago.
\begin{figure}[h]
\begin{center}
\includegraphics[width=0.47\columnwidth]{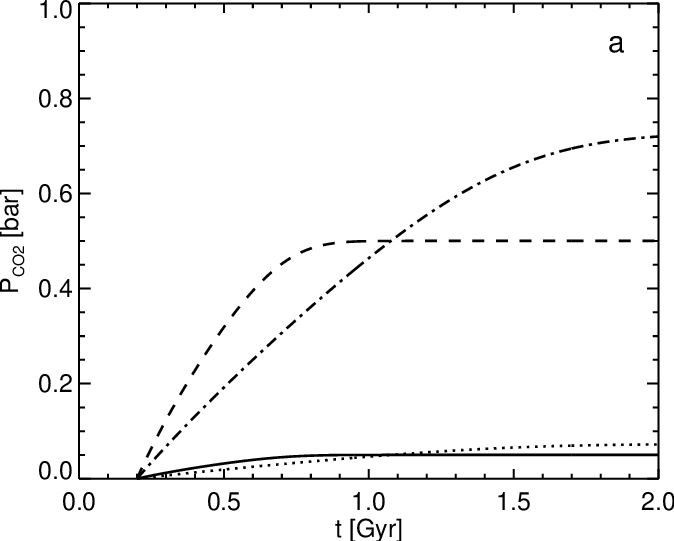}
\includegraphics[width=0.47\columnwidth]{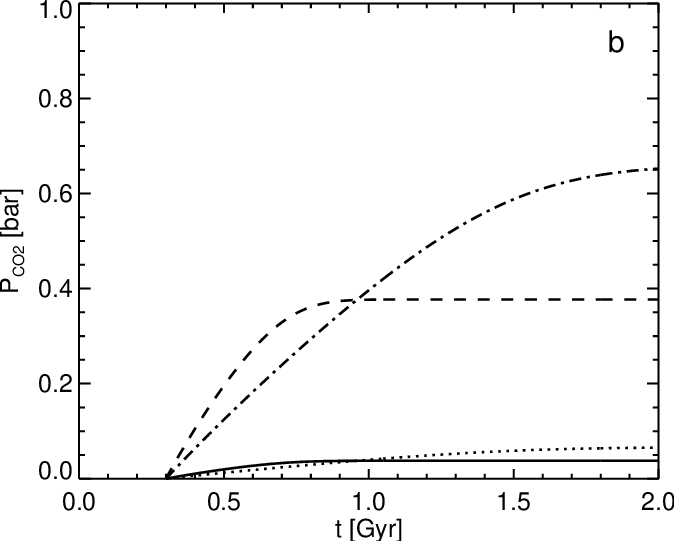}
\includegraphics[width=0.47\columnwidth]{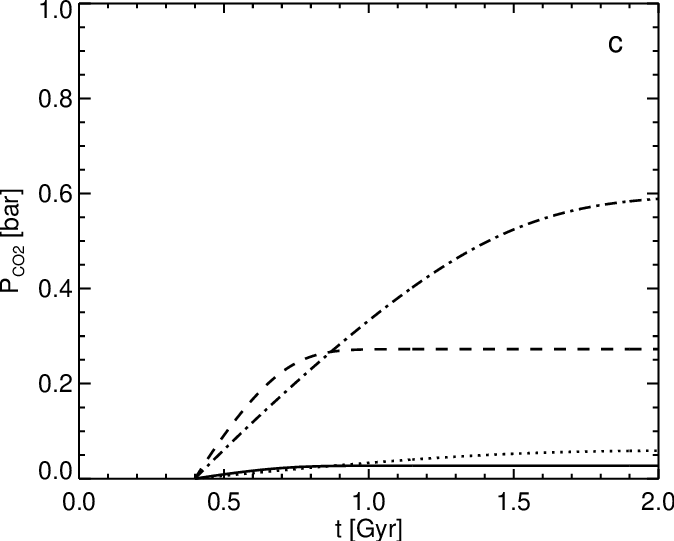}
\includegraphics[width=0.47\columnwidth]{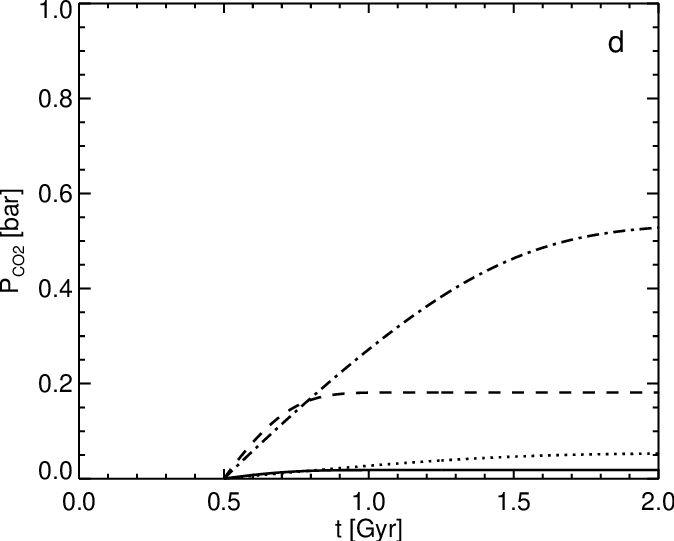}
\caption{CO$_2$ partial surface pressure as a function of time with the same assumptions as in Fig. 4 but for various onset times
for the build up of a secondary CO$_2$ atmosphere after total loss of the earlier outgassed CO$_2$ content.
Dashed lines: IW=1, Surface fraction of the melt channel $f_{\rm p}$=1; dashed-dotted lines: IW=1, $f_{\rm p}$=0.01; solid lines: IW=0, $f_{\rm p}$=1;
dotted lines IW=0 and $f_{\rm p}$=0.01. The onset for atmospheric growth of a secondary CO$_2$ atmosphere is assumed in
a: 4.3 Gyr (a), 4.2 Gyr (b), 4.1 Gyr (c), and 4 Gyr (d) ago.}
\end{center}
\end{figure}
\begin{figure}[h]
\begin{center}
\includegraphics[width=0.47\columnwidth]{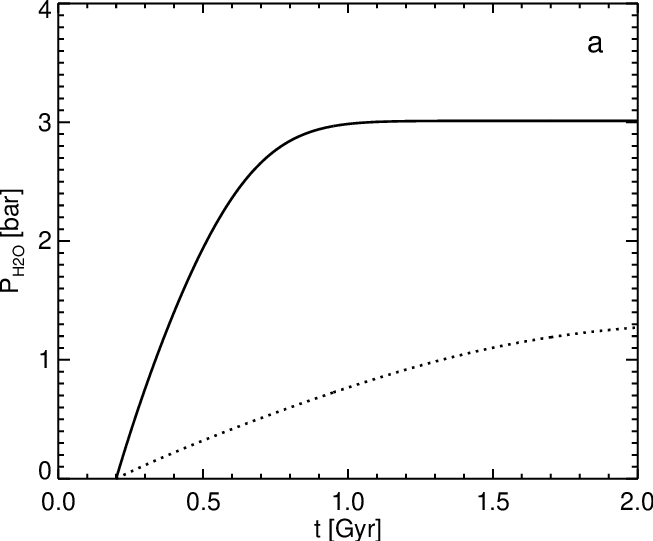}
\includegraphics[width=0.47\columnwidth]{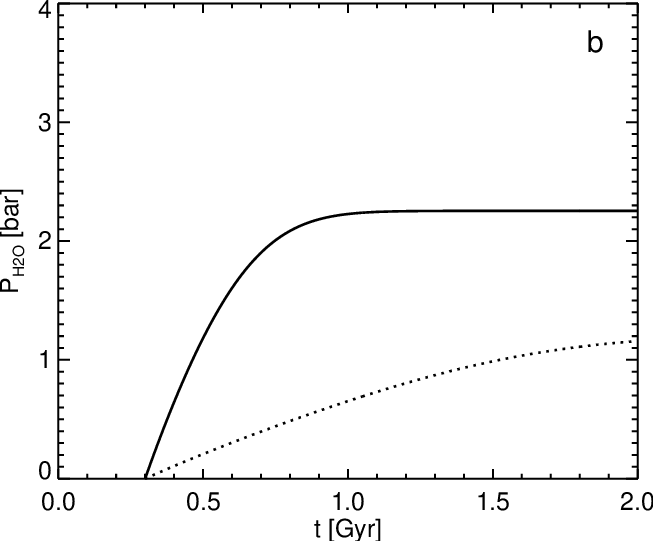}
\includegraphics[width=0.47\columnwidth]{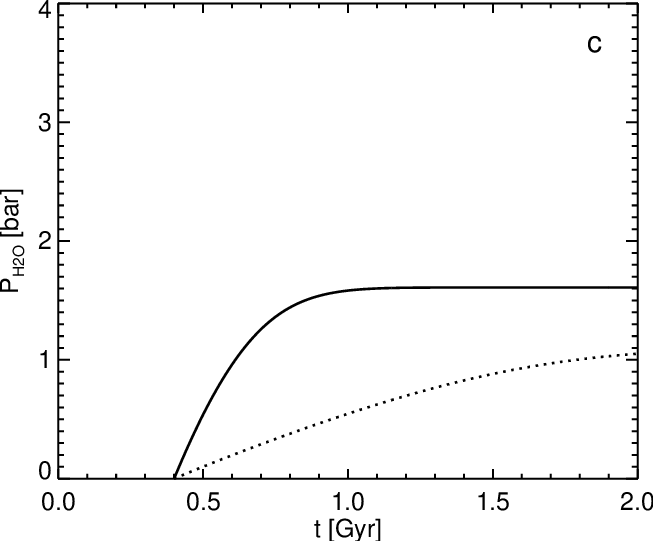}
\includegraphics[width=0.47\columnwidth]{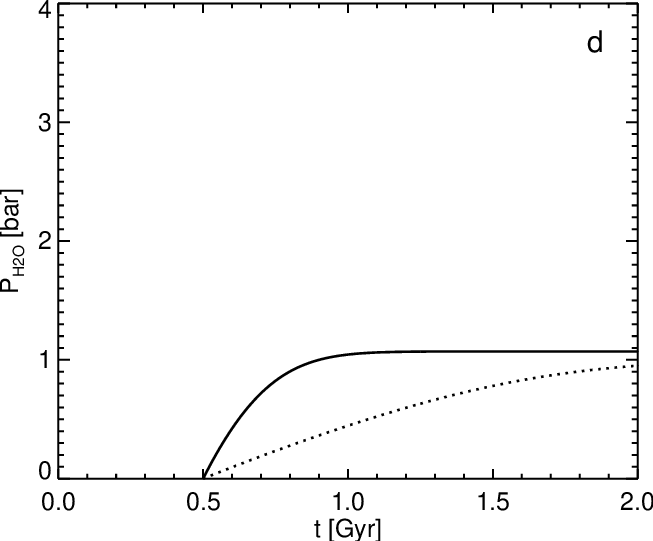}
\caption{Outgassed H$_2$O in units of bar as a function of time with the same assumptions as in Fig. 4 but for various onset times
where the outgassing flux exceeded the escape flux after the total loss of the earlier outgassed water content.
Solid lines: Surface fraction of the melt channel $f_{\rm p}$=1; dotted lines: $f_{\rm p}$=0.01. The bulk concentration of water in the
mantle is assumed to be 100 ppm and the outgassing efficiency is assumed to be 0.4.}
\end{center}
\end{figure}
Fig. 9 shows the possible scenarios for a build up of a secondary CO$_2$ atmosphere and Fig. 10 shows secondary outgassed
H$_2$O amounts in units of bar. Table 3 summarizes the accumulated outgassed CO$_2$ amount in units of bar after the
outgassed flux becomes more efficient compared to the escape flux for the same oxygen fugacity the related iron-wustite buffer (IW) and $f_{\rm p}$ scenarios shown in Fig. 9.
\begin{table}
\caption{Secondary outgassed CO$_2$ partial surface pressure in units of bar as function of iron-wustite (IW) buffer
and surface fraction of the melt channel ($f_{\rm p}$) (see Sect. 3 and Grott et al. 2011) and time $t$ after
Mars' origin in Myr when the outgassing flux is assumed larger than the escape flux.}
\begin{center}
\begin{tabular}{l|cccc}
\hline\noalign{\smallskip}
IW and $f_{\rm p} scenarios$& $t$=200 Myr &$t$=300 Myr  &$t$=400 Myr  &$t$=500 Myr \\\hline
IW=1; $f_{\rm p}$=0.01      & $\sim$0.7 bar & $\sim$0.65 bar & $\sim$0.6 bar & $\sim$0.55 bar \\
IW=1; $f_{\rm p}$=1         & $\sim$0.5 bar & $\sim$0.37 bar & $\sim$0.25 bar & $\sim$0.18 bar \\
IW=0; $f_{\rm p}$=0.01      & $\sim$0.07 bar & $\sim$0.12 bar & $\sim$0.11 bar & $\sim$0.1 bar \\
IW=0; $f_{\rm p}$=1         & $\sim$0.05 bar & $\sim$0.085 bar & $\sim$0.05 bar & $\sim$0.03 bar \\\hline

\noalign{\smallskip}
\end{tabular}
\end{center}
\end{table}
If we consider that the escape flux of CO$_2$ was less than that from the
interior 4.3 Gyr ago (Fig. 9a) a secondary CO$_2$ atmosphere of $\sim$0.7 bar could build up $\sim$4 Gyr ago, which is
about 100 times denser than the present atmosphere if one assumes
a global melt channel. By considering mantle plumes only, the outgassing would be finished about
4 Gyr ago and a CO$_2$ atmosphere of $\sim$0.5 bar could have been built up. If the escape
flux could balance the volcanic outgassing for longer times, depending if one assumes a global melt channel
or mantle plumes, only CO$_2$ atmospheres with lower upper densities of $\sim$0.2--0.4 bar could build up
$\sim$2.5--4 Gyr ago. For cases with low IW-buffers, the secondary CO$_2$ atmosphere would only
have a surface density between $\sim$50--100 mbar.

Low CO$_2$ surface pressure values would also agree with a study by Zahnle et al. (2008),
which is based on a photochemical CO$_2$ stability problem discussed by McElroy (1972), that a martian
CO$_2$ atmosphere much denser than several 100 mbar may be not be stable for a long time because
the CO$_2$ will be photochemically converted into CO over timescales of $\sim$0.1--1 Gyr.

Chevrier et al. (2007) investigated the geochemical conditions which prevailed
on the martian surface during the Noachian period by applying calculations of aqueous equilibria
of phyllosilicates. These authors found that Fe$^{3+}$-rich phyllosilicates most likely precipitated
under weakly acidic alkaline pH, which was a different environment compared to the following period
which was dominated by strong acid weathering that led to the observed martian sulphate deposits.
Chevrier et al. (2007) applied thermodynamic calculations which indicate that the oxidation state
of the martian surface should have been also high during early periods, which supports our results of
an early efficient escape of hydrogen.

However, equilibrium with carbonates implies that the
precipitation of phyllosilicates occurs at low CO$_2$ partial pressure. Thus, from these
considerations one would expect that the lower surface CO$_2$ pressure shown in Figs. 8
and Table 3 may have represented the martian atmosphere $\sim$4 Gyr ago. If geochemical processes
prevented the efficient formation of carbonates then a dense CO$_2$ atmosphere could not have
been responsible for a long-term greenhouse effect which is necessary to enable liquid water
to remain stable at the surface in the post-Noachian period. In such a case other greenhouse gases
such as CH$_4$, SO$_2$, H$_2$S, etc. (Kasting 1997; Johnson et al. 2008) would be needed
to solve the greenhouse-liquid water problem during the late Noachian.

Depending on the surface fraction of the melt channel $f_{\rm p}$ and the onset time of accumulation from
Fig. 10 and Table 4 one can see that the outgassed amount of H$_2$O by volcanos would correspond to values $\sim$1--3 bar,
that is a $\approx$20--60 m EGL.
One should also note that additionally to the secondary outgassed atmosphere significant amounts of water and
carbon may have been brought later by comets. According to Morbidelli et al. (2000), Lunine et al. (2003) the equivalent of
$\sim$0.1 terrestrial ocean of H$_2$O, that is a
$\sim$300 m deep EGL of water, could have been provided to Earth by comets during the few 100 Myr
following main accretion. The net budget of cometary impacts could hypothetically have also resulted in a net accretion
of several bars or even tens of bars of H$_2$O (and several 100 mbar of CO$_2$) coming from
infalling comets until the late Noachian or during the LHB as discussed in Sect. 6. A fraction of this impact delivered
H$_2$O, if all was not lost due to the high thermal escape rate (e.g. Tian et al. 2009) could be stored in the crust (Lasue et al. 2012; Niles 2012).
\begin{table}
\caption{Secondary outgassed H$_2$O partial pressure in units of bar with similar conditions and
assumptions as shown in Fig. 10.}
\begin{center}
\begin{tabular}{l|cccc}
\hline\noalign{\smallskip}
$f_{\rm p} scenarios$& $t$=200 Myr &$t$=300 Myr  &$t$=400 Myr  &$t$=500 Myr \\\hline
$f_{\rm p}$=1        & $\sim$3 bar & $\sim$2.25 bar & $\sim$1.6 bar & $\sim$1.1 bar \\
$f_{\rm p}$=0.01     & $\sim$1.3 bar & $\sim$1.2 bar & $\sim$1.1 bar & $\sim$1 bar \\\hline
\noalign{\smallskip}
\end{tabular}
\end{center}
\end{table}
Although it is not clear at the present how much CO$_2$ was in the martian atmosphere $\sim$4 Gyr ago,
the secondary outgassed and accumulated atmosphere was most likely denser
compared to the 7 mbar of today.
\section{Environmental Effects of the Late Heavy Bombardment}
Although the previous sections have shown that due to the high EUV flux of the young Sun, atmospheric escape models
do not favor a dense CO$_2$ atmosphere during the first Gyr, we now investigate possible effects of the late heavy bombardment (LHB) period. The ratios between critical mass vs.
tangent mass $n$ as discussed in Sect. 4 determines whether impacts  cause atmospheric erosion or if they are rather a source of volatile.
For investigating the upper limit of atmospheric erosion  related to the LHB,
we consider only  the lowest limit of $n=10$ and examine the number of impactors which are necessary to erode the martian atmosphere since the late Noachian
so that we end up with $\sim$7 mbar.

If the martian atmosphere can be eroded by impacts only, the numbers of impacts above the critical mass has to be higher
compared to the number computed in the exponentially decaying impact flux model.
We found from our calculations that depending on the initially assumed surface pressure
of $\sim0.1-1$ bar, one would need $\sim8000-15000$ impactors with masses equal or larger than $m_{\rm crit}$
to erode the martian atmosphere to a surface pressure of $\sim$7 mbars over the last 3.8 Gyr. However an exponentially decaying impactor flux model
gives only $\sim$86 for $\sim$0.1 bar and $\sim$30 for $\sim$1 bar for the number of impacts above $m_{crit}$ over this time period. These numbers are $\sim100-500$ times
smaller than the necessary number of large impacts. Therefore, by considering an exponentially decaying impact flux, it is unlikely
that the martian atmosphere with a surface pressure $\geq$0.1 bar was eroded by impacts during the past 3.8 Gyr.

The LHB, on the other hand, can provide the number of large impactors which is necessary to remove the atmosphere from Mars. After a period which
can most likely be characterized by a weak bombardment rate $\sim$3.9 Gyr ago, the planets experienced the LHB. The LHB was a cataclysmic
episode characterized by a high bombardment rate, during a time-span of $\sim50-300$ Myr. The Nice model  (e.g., Gomes et al., 2005)
which simulates the orbital evolution of the Solar System with slow migration of the giant outer planets, followed by a chaotic phase
of orbital evolution, yields an estimate of impactor mass distribution during this period. The impactor masses could be distributed as presented in Fig. 11 during the late
heavy bombardment period (data provided by Morbidelli, private communication). By investigating the best case for the erosion efficiency, we consider that the
largest impactor provides the first impact.

The maximum diameter of the impactors hitting Mars as a function of time can be compared with the critical impactor diameter $d_{\rm crit}$
that can erode the atmosphere. For an assumed initial surface pressure of $\sim$300 mbar, the critical mass $m_{\rm crit}$ and the
corresponding critical impactor diameter $d_{\rm crit}$  (with $\rho = 2000$ kg m$^{-3}$) for different values of $n = m_{\rm crit}/m_{\rm tan}$
can be calculated. By assuming $n=10$, from these calculations one obtains an upper limit for the amount of atmosphere
which can be eroded by impacts of $\sim$150 mbar over an intense bombardment period of $\sim$0.3 Gyr.
Lower values of $n$ can erode primordial atmospheres of $\sim$400 mbar ($n\approx$3) and even $\sim$1 bar in the case of $n$$\sim$1).
From these results one can see that impacts could have removed a major fraction of an accumulated secondary CO$_2$ atmosphere (see Fig. 6).
\begin{figure}[h]
\begin{center}
\includegraphics[width=0.75\columnwidth]{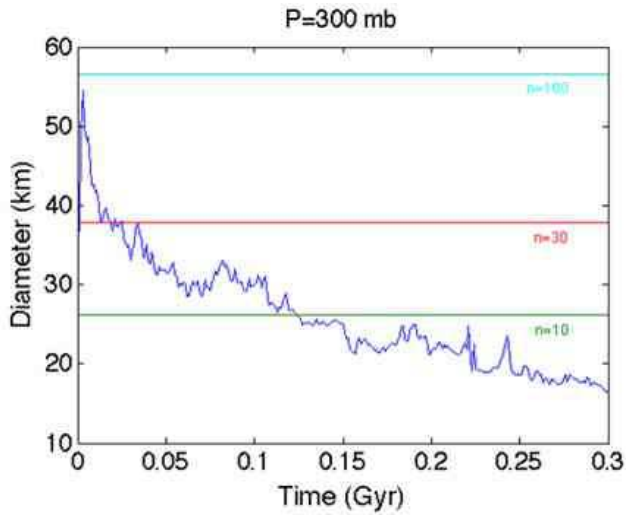}
\caption{Impactor diameters for the LHB for the best erosion case and for an initial pressure of $\sim$300 mbar.}
\end{center}
\end{figure}
The main problem with the impact studies remains to be the choice of parameter $n$. As discussed in Sect. 4 studies which
assume values for $n$ $\geq$30 deliver volatiles to the martian surface. Under this consideration, small $n$, and
hence atmospheric erosion due to impacts as discussed before, is questionable considering that recent hydrocode simulations
suggest at least an order of magnitude larger value of $n$ (Svetsov 2007; Pham et al. 2009; 2010; 2011). In such
a case the LHB would have accumulated volatiles additively to the secondary outgassed atmosphere.
By assuming that delivered CO$_2$ corresponds to $\sim$1 \% of the impactors this accumulation could result in an amount of
impact delivered CO$_2$ of $\sim$300 mbar. Thus, the H$_2$O which could have been brought to Mars by impacts (Levison et al. 2001)
especially during the LHB-period, where the solar EUV flux and related thermal escape processes were much lower compared to their early values,
could also be an important contribution to the planets present water inventory.
The cometary bombardment, is largely unconstrained but can deliver up to or even more then $\sim$5 bar of H$_2$O, that corresponds to a $\approx$130 m
deep GEL. These numbers should be considered as upper limits for the assumed total mass of the comets and
asteroids (7$\times$ $10^{22}$ g and 4$\times$ $10^{22}$ g, respectively) which may have fallen to Mars during
the LHB (data provided by Morbidelli 2009, private communication).

Geomorphological and geological evidence shows that liquid water flowed on the martian surface,
particularly in the Noachian period (Baker 2001;
Squyres and Knoll 2005). In order to have liquid water stable on the martian surface, CO$_2$ surface pressures of several bar
are necessary to obtain temperatures above freezing (Kasting 1991).
If one considers scattering of infrared radiation from CO$_2$-ice clouds (Forget and Pierrehumbert
1997) or additional greenhouse gases such as CH$_4$, SO$_2$ and H$_2$S,
which could have also been released by volcanism (Kasting 1997;
Johnson et al. 2008) this value can be achieved for $\sim$0.5--1.0 bar.
On the other hand aqueous solutions with lower melting points may have existed
(e.g., Fair\'{e}n, 2010, and references therein; M\"{o}hlmann 2012) making it possible that
Mars might have been ``cold-and-wet'' with average surface temperatures of $\sim$245 K
(Fair\'{e}n 2010; Gaidos and Marion 2003). Furthermore, water released by large impacts
during the LHB could also have liberated huge amounts of water so that
transient wet and warm conditions on the surface (Segura et al. 2002;
Toon et al. 2010) could have occurred.

However, if impacts delivered volatiles additionally to the secondary atmosphere
during the LHB period, this portion should have been lost partly to space during the Hesperian and Amazonian by various
nonthermal atmospheric escape processes and partly weathered out of the atmosphere into the
surface and ice.

\section{Escape and Surface Weathering of the Secondary Atmosphere Since the End of the Noachian}
From Mars Express ASPERA-3 ion escape data, Barabash et al. (2007) estimated the
fraction of CO$_2^+$ molecular ions lost to space since the end of the Noachian when the martian dynamo
stopped to work equivalent to a surface
pressure of about $\sim$0.2--4 mbar.
The present CO$_2^+$ escape rates are about two orders of magnitude lower
compared to the O loss and are on the order of $\sim$$8\times 10^{22}$ s$^{-1}$
(Barabash et al. 2007).

That direct escape of CO$_2^+$ ions from Mars was low is also in agreement with various MHD and
hybrid model results which yield an integrated CO$_2^+$ ion loss (IL) since the end
of the Noachian on the order of $\sim$0.8--100 mbar (e.g., Ma et al. 2004;
Modolo et al. 2005; Chassefi\`{e}re and Leblanc 2007; Lammer et al.
2008; Manning et al. 2010). Moreover from a recent study of Ma et al. (2007)
who calculated an escape rate of carbon of about 1.8 times larger at solar maximum
than at solar minimum which is in good agreement with the dependency of the escape ion rate
calculated in Fox et al. (2009) the estimated amount of CO$_2$ lost by ion loss since $\sim$4 Gyr
is most likely not in excess of $\sim$1 mbar (Chassefi\`{e}re and Leblanc 2011a). Thus,
from these studies we can consider that the realistic CO$_2^+$ molecular ion loss by pick up and outflow
through the martian tail in the theoretical range of
about 0.8--100 mbar given in Manning et al. (2010) should be considered closer to the
lower values.

On the other hand one should mention that all the previous ion escape models
did not use an accurately modeled neutral atmosphere and ionosphere which corresponds to
higher EUV fluxes expected before 2.5 Gyr ago. In such a case one can expect that more
carbon dioxide will be dissociated in the thermosphere so that it can be heated to higher
temperatures. As shown by Tian et al. (2009) a hotter thermosphere leads to an expansion of the
upper atmosphere and thus more extended coronae. In such a case the
solar wind interaction
area would be larger and one may expect higher ion loss rates too.

One should also note that solar wind induced forcing of Mars can also result in outflow and escape
of ionospheric ions. ASPERA-3 observations indicate that the replenishment of cold ionospheric ions
starts in the dayside at low altitudes at $\sim$300--800 km, where ions move at a low
velocity of $\sim$5--10 km s$^{-1}$ in the direction of the external magnetosheath flow (Lundin 2011).
The dominating energization and outflow process, applicable for the inner magnetosphere of Mars,
leads to outflow at energies of $\sim$5--20 eV. These energized ``cool'' ionospheric ions
can be picked up, accelerated by the current sheet, by waves and parallel electric
fields (Lundin 2011). The latter acceleration process can be observed above martian crustal magnetic
field regions. But even if we
assume that cold ionospheric ions may enhance the ion escape for carbon bearing species the
escape related to these processes most likely remains within the range given by Manning et al. (2010).

The Kelvin-Helmholtz (KH) plasma instability has also been regarded as a possible nonthermal
atmospheric loss process around unmagnetized planets since Pioneer Venus Orbiter observed detached
plasma structures, termed plasma clouds which contained ionospheric particles, downstream to the terminator in the
magnetosheath of the planet (Brace et al. 1982; Wolff et al. 1980). Around planets, magnetopauses or
ionopauses form boundaries with velocity shears, where the KH instability might be able to develop.
On their way along the boundary from the subsolar point to the terminator, waves of initially small
amplitudes grow and eventually form vortices in their nonlinear stage. When the vortex is able to detach,
it carries ionospheric particles away and thus can contribute to the loss of ions (Brace et al. 1982).

Amerstorfer et al. (2010) and M\"{o}stl et al. (2011) performed recent numerical simulations of the
KH instability with input parameters suitable for the boundary layers around unmagnetized planets.
Fig. 12 shows a time series of the normalized mass density at different times during one of their simulations. After the linear growth time of the instability,
a regular-structured vortex has evolved in the nonlinear stage. For this simulation, the density of the lower plasma layer is only ten times the density of the upper layer - a larger density jump stabilizes the boundary layer.
\begin{figure}[tb!]
\centering
\includegraphics[width=12cm]{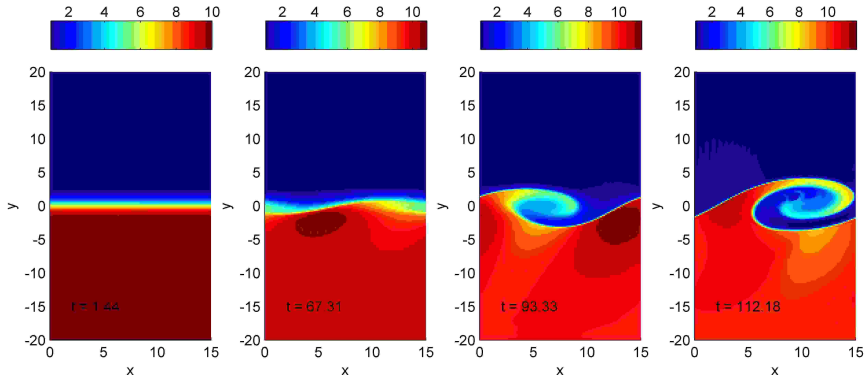}
\caption{Nonlinear evolution of the Kelvin-Helmholtz instability. The time series of the mass density is shown, from an MHD simulation with periodic boundary conditions in the
$x$-direction. The mass density changes from the upper to the lower plasma layer and exhibits an increase of up to
ten times (see the color code; blue: low density, red: high density). In the upper layer, the plasma flows from
left to right. In the lower layer, the plasma is at rest. Initially small perturbations of the boundary layer
separating the two plasma layers evolve into a KH vortex.}
\end{figure}
The results of M\"{o}stl et al. (2011) indicates that the martian ionopause should be stable with regard to the
KH instability due to the stabilizing effect of the large mass density of the ionosphere. However, the induced
magnetopause (Venus) or magnetic pile-up boundary (Mars) might be KH unstable during high solar activity.
For this boundary, the atmospheric loss of planetary ions might not be as severe as if the ionopause was the unstable boundary.
Thus this recent result indicate that
the loss due to the KH instability is not as significant as previously thought (Penz et al. 2004).
Furthermore, at the altitudes where one can expect that under extreme solar conditions plasma clouds may detach from the upper
atmosphere atomic oxygen ions should be the dominant species and CO$_2^+$ or CO$^+$ ions are most likely negligible constituents.

Atmospheric sputtering (SP) has been identified as an escape process of heavy
atoms from planetary bodies with low gravity such as Mars (Luhmann and Kozyra 1991).
Leblanc and Johnson (2002) studied sputter escape of CO$_2$ and CO from the
 martian atmosphere during the past 3.5 Gyr with a coupled test particle Monte Carlo molecular
dynamic model which considers collisions between photochemically produced
suprathermal atoms and background molecules for EUV fluxes which are 3 times and 6
times higher than that of today's Sun.
These authors obtained an escape of CO$_2$ caused
by sputtering since about 3.6--4 Gyr on the order of $\sim$50--60 mbar
(see also Chassefi\`{e}re and Leblanc 2007). More recently, it has been argued that the flux of
pick-up ions reimpacting Mars' atmosphere follows a logarithmic slope with the EUV flux of
$\sim$1.8 (Chassefi\`{e}re and Leblanc 2011a), much flatter than the value of $\sim$8 used
in Chassefi\`{e}re et al. (2007), yielding a cumulated sputtering escape rate since $\sim$4 Gyr
which gives a CO$_2$ loss $\leq$1 mbar.

One can see that the sputter loss was probably similar to that of ion erosion and for sure not
efficient enough to cause the loss of that hundreds of mbar of CO$_2$ could be lost. Further, we note that the sputtering
is a highly nonlinear process that depends on the EUV flux and the life time of the martian magnetic
dynamo (Dehant et al. 2007). Furthermore, it
was shown by Terada et al. (2009) that, due to the extreme solar wind atmosphere interaction caused by the young Sun before $\sim$4 Gyr ago,
a stronger induced magnetic field in the upper atmosphere could have decreased sputtering
during the transition period when the planet's intrinsic dynamo stopped working.

Besides ion escape and sputtering the loss of exothermal photochemically produced suprathermal
atoms such as O, C, N and H could have been more effective compared to both escape
processes discussed before. Exothermal processes such as dissociative recombination (DR)
of O$_2^+$, N$_2^+$ or CO$^+$ ions  produce neutral atoms in the ionosphere with higher
kinetic energy compared to the background atmosphere (e.g., Ip 1988; Nagy et al. 1990; Kim et al. 1998;
Lammer et al. 2000; Fox 2004; Fox and Ha\'{c} 2009; Krestyanikova and Shematovich 2006; Chaufray et al. 2007;
Valeille et al. 2009; Gr\"{o}ller et al. 2010; 2012). These newly created particles
collide with the cooler background gas, lose energy by collisions, transfer energy so that
a cold atom could become more energetic and finally a fraction of them reach the exobase
level and if their energy is larger than the escape energy they are lost from the
planet as neutrals. The production of these hot O and C atoms originating from DR
of O$_{2}^+$ and CO$^+$ molecular ions is also strongly related to the solar EUV flux,
and to an electron temperature dependent rate coefficient, where the total
energy of these newly produced O($^3$P, $^1$D), O($^3$P, $^1$S) and C($^3$P, $^1$D), O($^3$P, $^1$D, $^1$S) atoms
is a sum of their released energies ($\Delta E$) according to a DR reaction channel of kinetic and internal energy,
the latter being stored in molecules as vibrational and rotational energy.

Excited C atoms can also be produced via photo-dissociation (PD) of CO molecules
from ${\rm CO + h\nu \rightarrow C(^3P) + O(^3P) + \Delta E}$, with $\Delta E$
obtained as the difference of the photon energy and the energy which is needed to dissociate
the molecule and excite the newly produced atoms. Fox (2004) made a complete calculation of
all possible photochemical channels for the production of carbon escaping particles
and found that between 7.5$\times10^{23}$ C cm$^{-2}$ s$^{-1}$ to $\sim$4.5$\times10^{24}$ C cm$^{-2}$ s$^{-1}$
may escape from solar minimum to maximum conditions with photo-dissociation of CO being the most
efficient process.

If we apply a recently developed hot atom Monte Carlo model which
 selects the magnitude of the initial velocity of a newly produced hot particle randomly from the calculated velocity
distribution, which considers collisions that are based on the energy dependent total and differential cross sections
for elastic, inelastic and  quenching collisions and numerous cascaded hot particles
(Gr\"{o}ller et al. 2010; 2012) to present martian conditions, we obtain total escape rates for ``hot''
O atoms of $\sim$$9\times 10^{25}$ s$^{-1}$ and for hot C atoms $\sim$$2.7\times 10^{25}$ s$^{-1}$
during high solar activity conditions and $\sim$$3\times 10^{25}$ s$^{-1}$ for ``hot'' O atoms and
$\sim$$3\times 10^{24}$ s$^{-1}$  for hot C atoms during low solar wind conditions.
Our obtained hot O escape rates from present Mars are in good agreement with Chaufray et al. (2007) and
Valeille et al. (2009) and about a factor of 5 to 3 lower compared to results
from Fox and Ha\'{c} (2009) for high and low solar activity, respectively.

If we apply our model to the
thermosphere/ionosphere profiles modeled by Tian et al. (2009) during the earlier periods of martian history
preliminary estimates yield an integrated total CO$_2$ loss from today to 4 Gyr ago of $\leq$100 mbar.
A smaller value of $\leq10$ mbar has been proposed by Chassefi\`{e}re and Leblanc (2011a).

Table 5 compares expected max. and min. outgassed and accumulated CO$_2$ atmosphere in units of bar at $\sim$4 Gyr
and the integrated min. and max. CO$_2$ escape of various nonthermal atmospheric loss processes from observations
and models since that time.
Table 5 shows a very large dispersion of one or two orders of magnitude yielding a
range of total CO$_2$ loss from less than $\sim$10 mbar up to $\sim$100 mbar, which is definitely smaller
than the expected upper values of an outgassed and accumulated secondary CO$_2$ atmosphere.
However, a similar amount  of CO$_2$ or even a higher one could be stored
in the surface sinks which we discuss in the next section.
\begin{table}
\caption{Estimated min. and max. outgassed (VO: volcanic outgassing) CO$_2$ in units of bar $\sim$4 Gyr ago
and the expected min. and max. range of impact eroded (IE) or delivered (ID) atmosphere during the late heavy bombardment (LHB) period.
CO$_2$ escape of various atmospheric loss processes (IL: ion loss; KH: Kelvin Helmholtz instability triggered
ionospheric detached plasma clouds; SP: sputtering; DR: dissociative recombination; PD:
photo dissociation) from observations and models integrated since that time.}
\begin{center}
\begin{tabular}{l|c}
\hline\noalign{\smallskip}
   Sources and loss processes & CO$_2$ [bar]      \\\hline
   VO $\sim$4 Gyr ago, max.   & $\sim$0.2--0.5 bar   \\
   VO $\sim$4 Gyr ago, min.   & $\sim$0.05 bar        \\ \hline
   ID $\sim$ LHB              & $\sim$0--0.3 bar      \\
   IE $\sim$ LHB              & $\sim$0--0.15 bar  \\ \hline
   IL since $\sim$4 Gyr ago   & $\sim$0.001--0.1 bar\\
   KH since $\sim$4 Gyr ago   & $\sim$0.001 bar\\
   SP since $\sim$4 Gyr ago   & $\leq$0.001--0.05 bar \\
   DR since $\sim$4 Gyr ago   & $\sim$0.001--0.1 bar \\
   PD since $\sim$4 Gyr ago   & $\leq$0.005--0.05 bar \\
\noalign{\smallskip}
\end{tabular}
\end{center}
\end{table}
\section{Surface Sinks of CO$_2$ and H$_2$O}
Estimates of the total amount of CO$_2$ degassed into the martian atmosphere since accretion vary between $\sim$5--12 bar, while as shown in Sect. 5 most of
it was lost to space during the period of the young and active Sun. However, most of the strong EUV-powered atmospheric loss occurred most likely prior or during the early period of the Noachian and the Hellas impact ($\sim$4.0 Gyr). As it was discussed in the previous Sections the lost amount to space of CO$_2$ which accumulated from volcanic outgassing and/or was delivered by impacts during the LHB, is most likely
not higher than $\approx$100 mbar. Therefore, if Mars had indeed a denser CO$_2$ atmosphere $\sim$3.5--4 Gyr ago, most of it should be hidden below the planet's surface.
\subsection{Sequestration of CO$_2$ in Carbonates}
Possible sinks for this CO$_2$ atmosphere include loss to space, adsorption on the regolith, deposits of CO$_2$-ice, and deposits of carbonate minerals. The maximum CO$_2$ adsorbed on the regolith has been estimated to be on the order of $\sim$30--40 mbar and is likely to be less than that (Zent and Quinn 1995). Recent discoveries of buried CO$_2$-ice deposits in the polar regions indicate that modern CO$_2$-ice deposits could approach $\sim$20-30 mbar equivalent CO$_2$ (Phillips et al. 2011). Combining these sinks with the amount of CO$_2$ likely lost to space since the past $\sim$4 Gyr (see Table 5), it becomes clear that the majority of CO$_2$ that could be accounted for is most likely $<$150 mbar.

Therefore, the amount of CO$_2$ which is possibly stored as carbonate becomes the key to understanding the density of the ancient martian atmosphere at the end of the Noachian since it is the only sink that can accommodate a dense Noachian atmosphere.
Early studies modeling a dense atmosphere on Mars predicted abundant carbonates in the martian crust (Pollack et al. 1987), but after a decade of intense exploration of Mars from orbit and on the surface, abundant carbonate deposits have not been discovered (e.g., Bibring et al. 2007).
However, the detections of carbonates that have been made to date have revealed deposits that are either buried
or widely dispersed making detection difficult (Ehlmann et al. 2008; Michalski and Niles 2010;
Morris et al. 2010). Therefore, orbital detection of carbonates may not provide a complete view
of the carbonate crustal reservoir on Mars which may be larger than currently expected.
Nevertheless, information from martian meteorites, landers, rovers, and orbiters is now
available to construct a fairly consistent story of the carbonate reservoir on Mars.

The most direct evidence for carbonates on Mars comes from martian meteorites, many of which contain
carbonate minerals in trace abundances ($<$1\%) (Bridges et al. 2001). There are currently
40--50 known meteorites which are derived from 3--6 distinct sites on Mars (Eugster et al. 2002).
These rocks are derived from energetic ejection events from the martian surface which likely destroyed all but
the strongest igneous rocks providing a selection bias. Therefore it is clear that while martian meteorites
are invaluable samples of the martian surface, they likely do not provide an adequate sample by which to
judge the carbonate crustal reservoir of Mars. They do indicate that weathering and carbonate formation on
Mars has been active at least at very low levels throughout the Noachian, Hesperian, and Amazonian periods
(Gooding et al. 1988; Mittlefehldt 1994; Bridges et al. 2001; Niles et al. 2010). Localized deposits
of more concentrated carbonates have also been found in some of the oldest Noachian terrains on Mars.
These deposits, which have mostly been identified spectroscopically from orbit, are typically mixtures
of carbonate minerals with other phases including serpentine, olivine, smectite clays, and pyroxene
minerals (Ehlmann et al. 2008; Michalski and Niles 2010; Morris et al. 2010). They are also
buried underneath younger volcanic or ejecta deposits indicating ages that likely date to the
early to mid Noachian.

The carbonate abundance in these deposits ranges from $\sim$10--30\%
(Ehlmann et al. 2008; Michalski and Niles 2010; Morris  et al. 2010).
\begin{figure}[tb!]
\centering
\includegraphics[width=8cm]{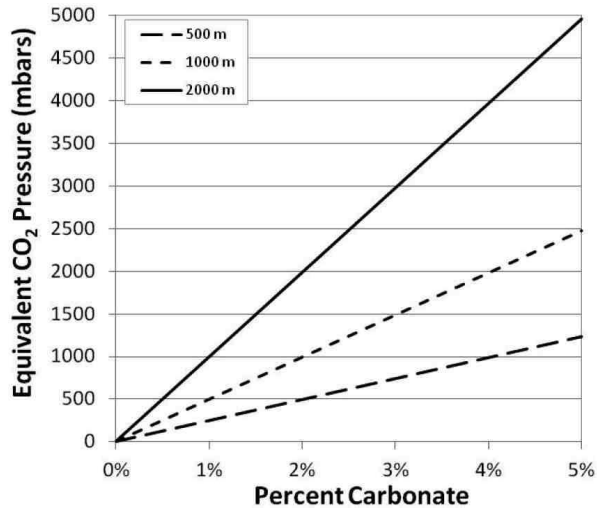}
\caption{Calculation of equivalent CO$_2$ atmospheric pressures based on carbonate reservoir size
calculated from volume percentage of carbonate and depth of carbonate mineralization.}
\end{figure}
In Gusev crater,
a carbonate deposit was identified by the Spirit rover using spectroscopic, chemical, and M\"{o}ssbauer data.
This deposit is also consisted of carbonate mixed with volcanoclastic materials, with carbonate abundances of $\sim$16--34\%.
Carbonates have also been detected in the dust and soils of Mars to be present at abundances between $\sim$2--5\%
(Bandfield et al. 2003). This is perhaps our best means for estimating the crustal reservoir of
carbonate on Mars as the dust on Mars is globally mixed (McSween and Keil 2000) and may be
representative of the average composition of the upper crust of the planet. However,
the dust is very fine grained and might be highly susceptible to weathering under current
martian conditions (Shaheen et al. 2010), and so might reflect an artificially elevated carbonate
abundance due to atmospheric or near surface weathering. Nevertheless, the carbonate content of the dust
can provide at least a hypothetical upper boundary for the carbonate content of the crust.

Fig. 13 shows the equivalent
atmospheric pressure for different crustal abundances of carbonate, assuming 5\% as an upper boundary
in the crust and crustal thickness up to 2 km. As one can see, a maximum amount of CO$_2$ which could hypothetically be stored in the crust
could be up to $\sim$5 bar with the assumption that the dust is a representative sample of the top 2 km.
Of course this estimate is for the total CO$_2$ reservoir in the martian crust, and only part of this or even
a tiny fraction of a carbonate reservoir formed since the Noachian.

It is difficult to determine how much of this carbonate reservoir can be attributed to the CO$_2$ presence during the
Noachian when many of the valley networks and phyllosilicates formed, but EUV-powered atmospheric escape was high.
The assumption that not much carbonates may have formed after the early Noachian is in agreement with the nature
of the carbonate deposits which have been discovered so far, as they all are dated at
the early to mid-Noachian or earlier (Ehlmann et al. 2008; Michalski and Niles 2010;
Morris et al. 2010).

Major outstanding issues remain in our understanding of the carbonate reservoir on Mars including:
\begin{itemize}
\item What was and is the total carbonate reservoir on Mars and how much carbonates are stored in the deep crust
($>$5 km)?
\item How efficient were environmental processes, which acted against carbonate formation?
\item How much carbonate has formed since the start of the Noachian and how is it distributed through time?
\item How efficiently are carbonates recycled back into the atmosphere on Mars and through what mechanisms?
\end{itemize}
\subsection{CH$_4$ as a Clue to an Active Long-Term Carbon Cycle}
During the last decade, CH$_4$ has been detected in Mars atmosphere by different instruments from both Earth
and spacecraft in orbit around Mars (Planetary Fourier Spectrometer (PFS) on Mars Express, Thermal Emission Spectrometer (TES)
on Mars Global Surveyor (MGS)) at an average
$\sim$10--20 ppb level (Krasnopolsky et al. 2004; Formisano et al. 2004; Mumma et al. 2009;
Fonti and Marzo 2010). One of the most striking characteristics of the observed CH$_4$ is its high temporal and
spatial variability, implying a lifetime of $\sim$200 days (Lef\`{e}vre and Forget 2009), much shorter than the
currently admitted value of $\sim$300 yr based on existing photochemical models (e.g. Krasnopolsky 2006). Such a large
discrepancy, together with the small signal to noise ratio of spacecraft data, led some scientists to consider the
detection of CH$_4$ as questionable. Concerning data acquired from Earth (Mumma et al. 2009), it has been
argued that the coincidence between martian and telluric lines could have led to an erroneous retrieval of the Mars
CH$_4$ abundance (Zahnle et al. 2011). Nevertheless, spacecraft data from MGS and Mars Express which now span $\sim$6
martian years (MY24--MY29) and Earth-based measurements show a globally consistent and reproducible seasonal behaviour
of the CH$_4$ mixing ratio with similar abundance levels and amplitudes of variation (see Fig. 1 in Chassefi\`{e}re and
Leblanc 2011b; Mousis et al. 2012, this issue), reasonably suggesting that CH$_4$ is effectively present in the martian atmosphere. All
existing observations show that CH$_4$ concentrations are spatially and temporally highly variable, and its lifetime has to
be on the order of $\sim$90 days. CH$_4$ could be oxidized in the superficial regolith layer through heterogeneous chemistry
processes involving gaseous oxidants like H$_2$O$_2$ and/or direct oxidation by ferric iron at the surface of grains
(Chassefi\`{e}re and Leblanc 2011a).

The calculated present CH$_4$ release flux, as deduced from present abundance measurements ($\sim$10 ppbv
CH$_4$ in a seasonal average, Geminale et al. 2008; 2011; Fonti and Marzo 2010) and estimates of the
CH$_4$ lifetime (200 days, Lef\`{e}vre and Forget, 2009) is on the order of $\sim$$1.0\times10^8$ cm$^{-2}$ s$^{-1}$.
Such a high flux is difficult to explain by an external (meteoritic) source and direct volcanism releasing only
trace amounts of reduced carbon can be similarly ruled out as a major source of CH$_4$ (Atreya et al. 2006).
According to these authors, hydrothermalism and/or biological activity may be at the origin of CH$_4$ on Mars.
It has been suggested by Dohm et al. (2008) that Mars is still internally active, with a potential for continuing
magmatic-driven activity (Hauber et al. 2011), including volcanism and hydrothermal activity, supporting a possible hydrothermal origin.

Hydrothermalism may result in the production of CH$_4$,
either by reduction of CO$_2$ through water-rock interaction at low oxygen fugacity (Lyons et al. 2006),
or through serpentinization followed by the conversion of H$_2$ in CH$_4$ (Oze and Sharma 2005).
Whatever the origin of CH$_4$,
assumed to take place in the crust below the water table, the fate of the produced CH$_4$ is to be transported upward
by ascending hydrothermal fluids and to be stored in the cryosphere in the form of CH$_4$ clathrate hydrates
(Chassefi\`{e}re and Leblanc 2011a; Mousis et al. 2012; this issue).

For how long has CH$_4$ been released at the present
rate to the atmosphere? A few arguments in favour of a long-term and relatively continuous phenomenon have been proposed:
\begin{itemize}
\item If CH$_4$ oxidation is at the origin of the suspected present redox imbalance between the H and O escape fluxes,
the release rate of methane averaged over the last 10$^3$ yr (photochemical lifetime of H$_2$) required to explain
the presumably small O escape rate, is close to the present release rate (Chassefi\`{e}re and Leblanc 2011b).
This suggests that methane could have been released at an average release rate similar to the present
one for at least a few thousand years.

\item As pointed out by Chassefi\`{e}re and Leblanc (2011b; 2011c), the quantity of carbon contained in the
superficial layer of CO$_2$-ice covering the permanent south polar cap is comparable to the amount of CH$_4$
which would be delivered to the atmosphere, then converted to CO$_2$, over a time interval of 3 Myr, that is the time since
the last obliquity transition period (Levrard et al. 2004). This small amount of CO$_2$, $\sim$1\% of the
atmospheric mass, would be at condensation equilibrium with larger amounts of CO$_2$-ice sequestered in the martian
south polar layered deposits, recently discovered by the Shallow Subsurface Radar (SHARAD) on Mars Reconnaissance Orbiter (MRO) (Phillips et al. 2010).
The thin superficial CO$_2$-ice layer covering the south polar cap could therefore originate in the hydrothermal CH$_4$,
as well as possibly some volcanic CO$_2$, released since the time of the transition. This suggests a
CH$_4$ release rate similar to the present value over the last few million or ten million years.

\item If, as may seem likely, CH$_4$ is released to the atmosphere from a CH$_4$ clathrate-rich cryosphere, the release rate of CH$_4$,
produced sporadically at depth by hydrothermal activity, is considerably smoothed over a time scale of the order of the lifetime of
CH$_4$ clathrate, that is the time for the cryosphere to fully sublimate to the atmosphere and possibly be recycled to the crust,
which may have occurred in the past. This time is estimated to be or the order of $\sim10^8$--$10^9$ yr
(Mousis et al. 2012, this issue), suggesting that CH$_4$ could have been continuously released to the atmosphere over geological time scales.
\end{itemize}
By scaling the CH$_4$ release rate on the level of hydrothermal activity, assumed to be proportional to the lava extrusion  rate as estimated from the geomorphological analysis of the surface (Greeley and Schneid 1991; Craddock and Greeley 2009), a cumulated amount of CO$_2$ resulting from CH$_4$ release since the Noachian of $\sim$2 bar has been estimated (Chassefi\`{e}re and Leblanc 2011a). Assuming that CH$_4$ release is due to serpentinization, a released amount larger than $\sim$0.4 bar is not consistent with the
present D/H ratio (see Sect. 3.1) in the martian atmosphere (Chassefi\`{e}re and Leblanc, 2011c). Up to $\sim$0.4 bar of CO$_2$ could therefore have been released in the form of CH$_4$, then oxidized. It should be noted that CH$_4$ outgassing is not an additional source
of carbon with respect to the CO$_2$ volcanic source described in the previous section. Either magmatic CO$_2$ is converted to CH$_4$ before
being released to the atmosphere through fluid-rock interaction in deep hydrothermal fluids (Lyons {\it et al.}, 2006),
or it precipitates in crustal carbonates, with further hydrothermal decomposition to CO$_2$, reduced
to CH$_4$ by the molecular hydrogen produced by serpentinization (Oze and Sharma 2005) and/or direct
thermodynamical equilibration (Lyons et al. 2006). Subsurface hydrothermal activity could be responsible for
both the release of CH$_4$ to the atmosphere, and the recycling of atmospheric CO$_2$ to the crust with
further precipitation of carbonates, recycled later to the atmosphere under reduced form through
hydrogeochemical processes (Chassefi\`{e}re and Leblanc 2011a).

Carbonate mineral deposits may occur
in subsurface hydrothermal systems from liquid water, rich in dissolved CO$_2$. Provided relevant subsurface
zones are not entirely sealed from the atmosphere, some CO$_2$ can be transferred from the atmosphere
to subsurface water reservoirs and further precipitate in carbonates, as observed on Earth and expected
to occur on Mars from geochemical modelling (Griffith and Shock 1995).
Such a long term carbon cycle, with a
progressive net removal of CO$_2$ from the atmosphere and subsequent carbonate deposition in the subsurface,
could explain the disappearance of a possible Noachian CO$_2$ atmosphere built by volcanism and late impacts.

Because a carbon atom may have been cycled several times through the crust since its release from the mantle by volcanism,
a cumulated CH$_4$ release rate up to $\sim$ 0.4 bar should not be interpreted as the content of a subsurface isolated reservoir.
It rather suggests that an efficient carbon cycle has been maintained by hydrothermal processes, probably until the early
Amazonian and possibly the present epoch, with a substantial fraction of the volcanic outgassed carbon being cycled
one or several times through crustal carbonates.
\subsection{Storage Capability of H$_2$O in the Martian Crust}
The present inventory of water on Mars is poorly constrained. The total water content of the two perennial polar caps
corresponds to a EGL  of H$_2$O of $\sim$16 m depth (Smith et al. 2001), and the ice deposits sequestered in the Dorsa
Argentea Formation (DAF), near the south polar cap, could have represented $\sim$15 m in the past (Head and Pratt 2001).
Nevertheless, only a fraction of the initial water could remain today in the DAF reservoir, corresponding to $\sim$5--7.5 m.
Other reservoirs, expected to have been active during late Amazonian, could be present in tropical and mid-latitude regions
(e.g., Watters et al. 2007; Holt et al. 2008;
Head and Marchant 2009), but they probably represent only a minor contribution to the global reservoir. The total inventory
of the known reservoir, including near-surface repositories that are distributed across middle to high latitudes, has been
estimated to correspond to a 35 m thick EGL (Christensen 2006). The megaregolith capacity is large, with up to $\sim$500 m
hypothetically trapped in the cryosphere, and hypothetically several additional hundreds of meters (up to $\sim$500 m) of ground
water surviving at depth below the cryosphere (Clifford et al. 2010). It has been suggested that most of ground ice
has been lost by sublimation at low latitudes, and that only small amounts of ground water would survive today (Grimm and Painter, 2009),
with therefore less water in the megaregolith. Carr (1987) suggested that a $\sim$500 m thick EGL of water has to be
required to explain the formation of outflow channels and most of this H$_2$O could be trapped today as water ice, and possibly
deep liquid water, in the subsurface.

Some of the water present on Mars may have reacted with minerals to form clay minerals and sulfates
(Bibring et al. 2006; Mustard et al. 2008; Ehlmann et al. 2012, this issue). The presence of these
hydrated minerals at the surface of Mars suggests that hydration processes have been active on Mars in the past.
They may have been formed, either at the surface of Mars during the Noachian, when liquid water was flowing
at the surface of the planet, or in the subsurface by aqueous alteration of subsurface rocks,
and possibly by impacts able to provide subsurface water to the impacted material (Bibring et al. 2006).
Existing geochemical model calculations show that hydrothermal hydration of martian crust is an efficient process
(Griffith and Shock 1997). Calculations have been made for a temperature in the range from 150--250$^{\circ}$ C.
The results do not depend much on the oxygen fugacity, and are similar for a moderately oxidized crust and for
a highly oxidized medium. The final conclusion of this study is that water storage via hydrous minerals can
account for $\sim$5 wt\% of crustal rocks. The capacity of the upper 10 km of the crust in storing water in the
form of hydrated minerals therefore hypothetically corresponds to a few hundreds meter depth EGL of H$_2$O, but the crustal content of
hydrated minerals is basically unknown and may be much lower than this upper limit.

The effectiveness and amplitude of aqueous alteration processes in Mars' crust are basically unknown.
SNC meteorites, originating in martian crust and mantle, provide information on crustal geochemical processes.
Formed of mantle material modified through interaction with crustal material during the upward migration of
lava through the crust, they have been shown to record a wide range of oxidation conditions (Wadhwa 2001).
Shergottites present a range of redox conditions from close to the IW-buffer up to that of the
quartz-fayalite-magnetite buffer, which can be interpreted as the result of oxidation of the crust by
a process such as aqueous alteration, through the oxidation of ferrous iron into ferric iron. Whereas
Nakhlites and Chassignites are oxidized, the ancient ALH 84001 is relatively reduced. These results
suggest that the silicates of SNC meteorites originate in a water-depleted martian mantle. These silicates
would have been oxidized through assimilation of oxidized crustal material. In order to produce such an
oxidation, a significant proportion of crustal material (10--30\%) with a high Fe$_3$$^+$/Fe ratio ($\sim$50\%)
must have been mixed with mantle material (Herd et al. 2002). ALH84001, because it has crystallized
at the end of the Noachian $\sim$3.9 Gyr ago, could have not been mixed with oxidized crust material.

The occurrence in the crust of aqueous alteration processes, as proved by oxidation processes recorded in
SNC meteorites, suggests that the conditions for an efficient hydration of crustal rocks may have been
met at some places and times in the past. Such hydrated minerals have been found in Nakhlites.
A particular hydration process occurring in Earth's crust is serpentinization, which generates H$_2$
from the reaction of water with ferrous iron derived from minerals, primarily olivine and pyroxene (McCollom and Back 2009).
In the reaction, ferrous iron is oxidized by the water to ferric iron, which typically precipitates as magnetite,
while hydrogen from water is reduced to H$_2$. Iron oxidation is accompanied by the storage of a large number
of water molecules in serpentine, an hydrated mineral which has been recently observed by the
Compact Reconnaissance Imaging Spectrometer for Mars (CRISM) on
MRO in and around the Nili Fossae region (Ehlmann et al. 2009). Serpentinization occurring in crustal hydrothermal
systems is a plausible process at
the origin of the methane observed in the martian atmosphere (Oze and Sharma 2007). Based on an analysis
of the present Mars' D/H ratio, it has been suggested that a water GEL of up to $\sim$300-400 m depth could
have been stored in crustal serpentine since the late Noachian due to hydrothermalism triggered by
magmatic activity (Chassefi\`{e}re and Leblanc 2011). Massive serpentinization of the southern crust
could have been at the origin of both the crustal dichotomy and the strong remanent magnetic field of old
southern terrains (Quesnel et al. 2009).

Although there is no direct observational evidence of active
hydration processes in Mars' crust, aqueous alteration processes, e.g. serpentinization, are
potentially able to store several hundreds of meters of H$_2$O in crustal hydrates. This
amount is comparable to the estimated value of the water ice content of the cryosphere.
Depending on the efficiencies of the various atmospheric escape processes and volatile delivery by impacts,
several hundreds of meters of water could hypothetically be trapped in the subsurface in the form of
H$_2$O-ice and/or hydrated minerals, and possibly liquid water below the cryosphere. All these reservoirs
could exchange with each other, as well as with the atmosphere and polar caps (Lasue et al. 2012), at the occasion of
magmatic and hydrothermal events, and could contain hypothetically up to $\sim$1 km thick EGL of water.
\begin{figure}[h]
\begin{center}
\includegraphics[width=11cm]{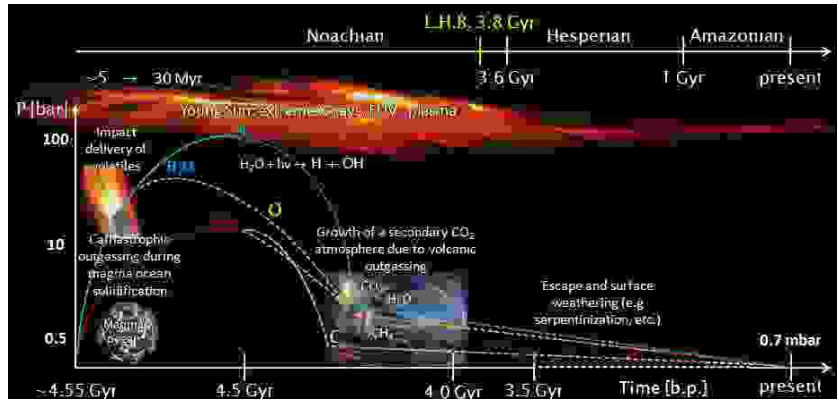}
\caption{Illustration of Mars' atmosphere evolution after the outgassing of volatiles and fast growth of a dense water vapour
dominated H$_2$O/CO$_2$ atmosphere during and after the magma ocean solidification process. blue lines: water and hydrogen;
yellow lines: oxygen; red lines CO$_2$. Solid lines: Outgassed or delivered species; dashed-lines: escaping species.
A complex interplay between the
young Sun's EUV activity, impacts and atmospheric escape processes during the first $\sim$500 Myr kept early Mars most of the time
cool and dry. After the EUV flux of the young Sun decreased at $\sim$4--4.3 Gyr ago, volcanic outgassing and
impacts during the LHB could have resulted in the build-up of a denser and slightly warmer and wetter CO$_2$ atmosphere
of $\leq$1 bar. If the secondary CO$_2$ atmosphere after the Noachian had a surface pressure which was larger than $\sim$100 mbar,
only a minor fraction escaped to space during the Hesperian and Amazonian and most of it should be stored
so far undiscovered in the subsurface. CO$_2$ can be released from surface reservoirs during periods of climate change or
impacts (non-linear dashed red line from the late Noachian until present) and may have modified the atmospheric
surface pressure several times during the planet's history.}
\end{center}
\end{figure}
\section{Conclusions}
The latest hypotheses on the formation, outgassing and evolution of the martian atmosphere from
the early Noachian up to the present time as illustrated in Fig. 14 have been discussed.
Depending on the captured nebula gas and/or outgassed amounts of volatiles, we show that due to the high EUV flux of the young Sun the planet's hydrogen-rich
protoatmosphere was lost via hydrodynamic escape of atomic hydrogen which
could have dragged heavier atoms such as C and O during the first tens or hundreds of Myr after Mars'
finished its accretion. The early
Noachian impacts may have kept the protoatmosphere in vapor form and may not have
much contributed to atmospheric growth because the delivered volatiles would have
also escaped. After Mars lost its protoatmosphere the atmospheric escape rates were
most likely balanced with a secondary outgassed
atmosphere and delivered volatiles by impacts until the activity of the young Sun
decreased so that the atmospheric sources could dominate over the losses. Depending on assumptions
related to geochemical conditions such as the pH of the early martian environment,
the global melt channel, melting in mantle plumes, oxygen fugacities and IW-buffers
in combination with atmospheric escape and impact delivery during the late heavy bombardment
Mars may have built up a secondary CO$_2$ atmosphere which was $<$1 bar
and accumulated a water inventory equivalent to $<$10 bar at $\sim$3.5--4 Gyr ago.
By reviewing the latest observations and model studies on the escape of the martian CO$_2$ atmosphere
during the Hesperian and Amazonian epochs we expect that the planet may have
lost most likely less than $\sim$150 mbar of CO$_2$. If a CO$_2$ atmosphere of several 100 mbar was indeed
present at the end of the Noachian,
most of it should have been removed to the crust by sequestration in carbonate rocks and partially recycled to the atmosphere under
reduced and/or oxidized form. The water contained in a several 10 m or even 100 m deep EGL at the Noachian
could have been trapped hypothetically in the crust in the form of H$_2$O-ice, hydrated minerals,
and possibly liquid water under the cryosphere. On the other hand, if there are not enough hidden
CO$_2$ deposits under the planet's surface, then other greenhouse gases are necessary for the explanation
of standing bodies of liquid water on the planet's surface $\geq$3.5--4 Gyr ago.
\begin{acknowledgements}
D. Breuer, E. Chassefi\`{e}re, M. Grott, H. Gr\"{o}ller, E. Hauber, H. Lammer, P. Odert and A. Morschhauser acknowledges
support from the Helmholtz Alliance project ``Planetary Evolution and Life''. E. Chassefi\`{e}re acknowledges support from
CNRS EPOV interdisciplinary program.
H. Lammer acknowledge the support by the FWF NFN project S116 ``Pathways to Habitability:
From Disks to Active Stars, Planets and Life'', and the related FWF NFN subproject,
S116607-N16 ``Particle/Radiative Interactions with Upper Atmospheres of Planetary Bodies Under Extreme Stellar Conditions''.
H. Gr\"{o}ller and H. Lammer acknowledges also support from the Austrian FWF project P24247-N16 ``Modelling of non-thermal processes
in early upper atomospheres exposed to extreme young Sun conditions'' and
support from the joined Russian-Austrian project under the
RFBR grant 09-02-91002-215-ANF-a and the Austrian Science Fund (FWF) grant I199-N16.
P. Odert was supported via the FWF project grant P19446-N16
and the research by U. M\"{o}stl was funded by the FWF project grant P21051-N16.
O. Karatekin thanks A. Morbidelli for the discussions related to impact studies and the LHB; O. Karatekin, V. Dehant and
L. B. S. Pham acknowledges the support of Belgian PRODEX program managed by the ESA
in collaboration with the BELSPO. O. Mousis acknowledges support from CNES. P. Niles
acknowledges support from NASA Johnson Space Center and the Mars Fundamental Research Program.
The authors also thank ISSI for hosting the conference and the Europlanet RI-FP7
project and its related Science Networking (Na2) working groups. Finally, the authors
thank guest editor M. Toplis and two anonymous referees for their suggestions and recommendations which helped to improve the article.
\end{acknowledgements}

\end{document}